%% file: main.tex
\documentclass[a4paper,12pt,english]{article}

\usepackage[authoryear,round]{natbib}
\setcitestyle{aysep={}} 

\newcommand{\bcitep}[1]{\textcolor{blue}{\citep{#1}}}
\usepackage{mathrsfs}
\usepackage{amsmath}
\usepackage{amssymb} 
\usepackage{url}
\usepackage{color}
\usepackage{graphicx}
\usepackage{multirow}
\usepackage{multicol}
\usepackage{subcaption}
\usepackage{placeins}
\usepackage{siunitx}
\usepackage{threeparttable}
\usepackage{rotating}
\usepackage{booktabs}
\usepackage{threeparttable}
\usepackage{lscape}
\usepackage{dsfont}
\usepackage{comment}
\usepackage{xcolor}

\newcommand{\blueeqref}[1]{\textcolor{blue}{\eqref{#1}}}
\usepackage{amsthm}
\newtheorem{proposition}{Proposition}
\newtheorem{theorem}{Theorem}
\newtheorem{lemma}{Lemma}

\newcommand{\F}{\mathscr{F}}

\graphicspath{{figuras/}}

\title{Threshold Generalized Autoregressive Moving Average Models}
\author{
  Katerine Zuniga Lastra\thanks{Institute of mathematics, statistics, and computer science, USP - zunigalastrakaterine@ime.usp.br},
  Airlane Pereira Alencar\thanks{Institute of mathematics, statistics, and computer science, USP - lanealencar@usp.br}
}

\date{\today}

\begin{document}

\maketitle
\begin{abstract}
Non-Gaussian time series often exhibit nonlinear dynamics and changes in dependence structures that may not be adequately captured by conventional single-regime models. To address this limitation, we introduce a class of threshold generalized autoregressive moving-average (TGARMA) models that combines observation-driven GARMA dynamics with threshold-based regime switching. The proposed framework accommodates multiple regimes, determined by a lagged endogenous or exogenous threshold variable, and allows regression coefficients, autoregressive and moving-average orders and coefficients, and, where applicable, dispersion parameters to vary across regimes. 
Parameters are estimated by maximum partial likelihood, while model selection is conducted using information criteria. We examine model admissibility and stability, assess model adequacy through residual diagnostics, and conduct a Monte Carlo simulation study to investigate the finite-sample performance of the estimators. The proposed methodology is illustrated using monthly counts of hospitalizations due to respiratory diseases among older adults in São Paulo, Brazil. In this application, the TGARMA model outperforms its single-regime GARMA counterpart in terms of in-sample mean squared error, highlighting the potential benefits of incorporating regime-dependent dynamics into the analysis of non-Gaussian time series.

\end{abstract}

\section{Introduction}
Time series analysis is fundamental to modern statistics, with applications spanning economics, finance, environmental sciences, epidemiology, engineering, and beyond. Classical autoregressive moving-average (ARMA) models and their integrated extensions (ARIMA) have long been the standard tools for modeling serial dependence in Gaussian processes \citep{box2015time,brockwell2009time,shumway2016time}. However, many contemporary applications involve responses that are intrinsically non-Gaussian---counts, binary outcomes, proportions, rates, durations, or positive continuous variables with skewed distributions. Applying Gaussian models to such data leads to inefficient estimation, unreliable predictive intervals, and invalid probabilistic interpretations.

The generalized linear model (GLM) framework \bcitep{nelder1972,McCullagh} provided a unified approach for independent non-Gaussian responses through exponential-family distributions. Extending this framework to dependent observations gave rise to observation-driven dynamic models, including the quasi-likelihood approach of \bcitep{zeger1988}, the conditional likelihood theory of \bcitep{li1994}, and the partial likelihood framework of \bcitep{Fokianos2004}. Among these, the Generalized Autoregressive Moving Average (GARMA) model \bcitep{benjamin2003} has become particularly influential. GARMA models generalize classical ARMA by specifying the conditional mean through a link function, while allowing autoregressive and moving-average terms in the linear predictor:
\[
g(\mu_t)=\mathbf{x}_t^\top \boldsymbol{\beta}+\sum_{i=1}^p \phi_i\{g(Y_{t-i})-\mathbf{x}_{t-i}^\top\boldsymbol{\beta}\}+\sum_{j=1}^q \theta_j\{g(Y_{t-j})-\eta_{t-j}\},
\]
where \(\mu_t=\mathbb{E}(Y_t\mid\F_{t-1})\), \(g\) is a link function, and \(\F_{t-1}\) denotes the past history. This formulation preserves the interpretability of ARMA components while accommodating Poisson, binomial, gamma, inverse Gaussian, and other exponential-family distributions. Subsequent extensions have addressed overdispersion and zero-inflation \bcitep{wagh2018zero,andrade2024zero,fadel2026}, bounded responses \bcitep{Bayers,maior2018symarma,bayer2023inflated}, alternative count distributions \bcitep{melo2020,sales2022berg}, process monitoring \bcitep{albarracin2018cusum}, and multivariate settings \bcitep{ribeiro2025bivariate}. Inference is typically based on conditional or partial likelihood, with asymptotic arguments relying on martingale estimating equations and conditional information \bcitep{andersen1970,Fokianos2004}; Bayesian approaches have also gained traction \bcitep{silveira2015,pala2023,lastra2025}.

Despite these advances, standard single-regime GARMA formulations keep the regression coefficients, autoregressive parameters, and moving-average terms fixed over time. This restriction can be inadequate when structural breaks or regime-dependent dynamics induce different behavior across states. For example, monthly insurance claims may exhibit low persistence during stable economic periods but strong autocorrelation following financial shocks, while the claims distribution remains overdispersed Poisson throughout. A single-regime GARMA model cannot represent such regime-switching, yet a Gaussian threshold model would misrepresent the discrete nature of the data. 
Threshold models offer a natural solution for regime-switching. Introduced by \bcitep{tong1978,tong1983,tong1990}, threshold autoregressive (TAR) models partition the state space according to a lagged threshold variable, allowing distinct autoregressive dynamics in each regime. This approach has been extensively developed for Gaussian and approximately Gaussian processes, including extensions to moving-average components (TARMA) \bcitep{goracci2019threshold,chan2019ergodicity}, with rigorous inferential theory \bcitep{chan1990likelihood,chan1993consistency,tong2011threshold,hansen2011threshold,zhang2023estimation,goracci2021empirical,goracci2023testing,goracci2025robust}. However, threshold methodology for non-Gaussian observation-driven models is much less developed, with existing contributions largely focused on specific distributions, such as count time series \bcitep{chen1995bayesian, pan2017bayesian, yang2022mcmc}. This leaves room for a general threshold framework that retains the distributional flexibility of exponential-family GARMA models.
Developing such a framework presents substantial challenges. The threshold parameter is not identified under the null hypothesis, complicating standard inference. The observation-driven dynamics interact with regime switching in ways that affect stationarity and ergodicity. Moreover, the likelihood function is discontinuous in the threshold parameter, requiring specialized asymptotic theory. These challenges have prevented a general solution.

This paper introduces the Threshold Generalized Autoregressive Moving Average (TGARMA) model. The proposed model, formally defined in Section~\ref{sec:proposed_model}, extends the GARMA framework by allowing regression coefficients, autoregressive parameters, moving-average terms and, when appropriate, dispersion components to vary across regimes determined by a lagged threshold variable. By retaining an exponential-family conditional distribution, the formulation covers common non-Gaussian responses such as counts, binary outcomes, proportions and positive continuous data, subject to the usual support and link-function restrictions. 

The contributions of this paper are as follows:
\begin{enumerate}
    \item We propose the TGARMA class, allowing dynamic, regression and dispersion-related quantities to vary across threshold regimes while retaining an observation-driven exponential family conditional distribution.
	\item We develop a partial maximum likelihood approach for estimating the continuous model parameters, together with an information criterion based grid search for selecting threshold values, threshold delays, and regime specific autoregressive and moving average orders.
    \item We derive the partial score function and the conditional information matrix, and present covariance estimators based on the observed Hessian and the sandwich formulation involving the outer product of gradients (OPG). We present an asymptotic normality result for the continuous parameter estimator within a martingale based partial likelihood framework.
    
	\item We clarify the role of threshold parameters. They are selected by profile likelihood over an admissible grid, but the regular Wald theory developed here applies only to the continuous parameters conditional on the selected threshold configuration.
    
	\item We discuss structural properties, admissibility conditions and stability diagnostics, emphasizing that stationarity and ergodicity of the full threshold recursion do not follow automatically from within-regime ARMA restrictions.
    
    \item We provide practical implementation guidelines, including PMLE for the continuous parameters combined with profile-likelihood estimation of the threshold vector and delay parameter, model selection using AIC and BIC \bcitep{akaike1974,schwarz1978}, residual diagnostics based on martingale differences and randomized quantile residuals \bcitep{dunn1996randomized}, and multi-step forecasting procedures.
	\item We illustrate the proposed methodology using a real-world monthly hospitalization count time series, demonstrating the practical applicability of the TGARMA framework.
\end{enumerate}

The remainder of the paper is organized as follows. Section~\ref{sec:proposed_model} defines the TGARMA model and discusses its probabilistic structure. Section~\ref{sec:estimation} develops partial maximum likelihood estimation, including the score, conditional information matrix and large-sample result for the continuous parameters. Section~\ref{sec:asymptotics} presents Wald-type inference for the continuous parameter vector. Section~\ref{sec:properties} discusses structural properties, admissibility and stability diagnostics. Section~\ref{sec:implementation} covers implementation aspects, including model selection, diagnostics, and forecasting. Section~\ref{sec:simulation} describes the simulation, and Section~\ref{sec:application} presents the real-data application. Section~\ref{sec:conclusion} concludes with a discussion and future research directions.

\section{Threshold Generalized Autoregressive Moving Average Models (TGARMA)}
\label{sec:proposed_model}
In this section, we introduce the general formulation of Threshold GARMA (TGARMA) models, which combine the generalized linear model and the autoregressive and moving average dynamics of GARMA models and a threshold-induced regime-switching mechanism.

\subsection{The proposed model}
\label{subsec:proposed_model}

Let $\{Y_t\}_{t \in \mathbb{Z}}$ be the stochastic process of interest and let $\{\mathbf{X}_t\}_{t \in \mathbb{Z}} \in \mathcal{X} \subset \mathbb{R}^{k+1}$ be a sequence of $(k+1)$-dimensional exogenous covariate vectors, possibly random, where $\mathbf{x}_t = (1, x_{t1}, \ldots, x_{tk})^\top$ includes a constant term. 
Let $R \in \mathbb{N}$, with $R \ge 2$, be the number of regimes, $d \in \mathcal{D} \subset \mathbb{N}$ the delay parameter, and $\mathbf{r} = (r_1, \dots, r_{R-1})^\top \in \mathcal{R} \subset \mathbb{R}^{R-1}$ the threshold vector, where  \(  \mathcal{R} = \{(u_1,\dots,u_{R-1})^\top \in \mathbb{R}^{R-1}: u_1 < u_2 < \cdots < u_{R-1}\}.\) The thresholds define a partition of the real line into $R$ distinct regimes indexed by $s \in \{1,\dots,R\}$ based on a lagged transition variable. Denoting the transition variable by $\{Z_t\}_{t \in \mathbb{Z}}$, the active regime at time $t$ is given by  
\begin{equation}
	S_t(\mathbf{r},d) = 
	\begin{cases}
		1, & Z_{t-d} \le r_1,\\
		2, & r_1 < Z_{t-d} \le r_2,\\
		\vdots & \vdots\\
		R, & Z_{t-d} > r_{R-1},
	\end{cases} \qquad t = m,\dots,T,
	\label{eq:regime_threshold_geral}
\end{equation}
with $S_t(\mathbf{r},d) \in \{1,\dots,R\}$ and
$m = \max\{p_1,\dots,p_R,q_1,\dots,q_R,d\}$, where $(p_s,q_s)$ are the autogressive and moving average orders for the s-th regime. The value of $m$ ensures that all
lagged quantities entering the recursion are available after conditioning on
the first $m$ observations. The sequence $\{S_t(\mathbf r,d)\}$ is random
whenever the transition variable $Z_t$ is random.


Define  $\F_t = \sigma(\mathbf{x}_t,\mathbf{x}_{t-1},\dots, Y_t,Y_{t-1},\dots, Z_t, Z_{t-1}$ the sigma algebra generated by all available information until time t;
then $\F_{t-1}$ represents the information available up to time $t-1$. 
In the usual selt-excited threshold model ?citeSETAR?,
$Z_{t-d}=Y_{t-d}$.
The covariate process may contain deterministic, stochastic, time-dependent components, or any combination thereof \bcitep{pumi2025matsuoka}. When stochastic covariates are present, inference is conducted conditionally on the observed covariate path \bcitep{slud1994partial, kedem2002regression, Fokianos2004}, and $\mathbf x_t$ is assumed to be predictable with respect to $\F_t$, which will be used in the partial likelihood. 
Care is required when stochastic covariates exhibit stochastic trends, since unrelated nonstationary series may produce spurious regression effects \bcitep{granger1974spurious, phillips1986understanding}; in such cases, detrending, differencing, or cointegration-based modelling restrictions may be needed \bcitep{engle1987cointegration}.

Following \cite{benjamin2003}, for each regime $S_t(\mathbf r,d)=s$ the conditional distribution of $Y_t$ given $\F_{t-1}$  belongs to the exponential family in canonical form. Hence, the conditional density or probability mass function can be written as
\begin{equation}
	f\bigl(y_t\mid\F_{t-1}, S_t(\mathbf r,d)=s\bigr)=
	\exp\left\{\frac{y_t\,\omega_t^{(s)} -b\bigl(\omega_t^{(s)}\bigr)}{\varphi^{(s)}}+c\bigl(y_t,\varphi^{(s)}\bigr)
	\right\},
	\label{eq:ef_geral}
\end{equation}
where $\omega_t^{(s)} \in \Omega_s \subset \mathbb{R}$ is the canonical parameter associated with regime $s$, $\varphi^{(s)} \in \Phi \subset \mathbb{R}^+$ is the regime-specific dispersion parameter, and $b(\cdot)$ and $c(\cdot,\cdot)$ are known functions that define the particular distribution. The conditional mean and variance are $\mu_t^{(s)}
= \mathbb{E}\bigl(Y_t \mid \F_{t-1}, S_t(\mathbf r,d)=s\bigr)
= b'\bigl(\omega_t^{(s)}\bigr)$ and $\operatorname{Var}\bigl(Y_t \mid \F_{t-1}, S_t(\mathbf r,d)=s\bigr)= b''\bigl(\omega_t^{(s)}\bigr)\,\varphi^{(s)} = \varphi^{(s)} V(\mu_t^{(s)})$, with $V(\mu_t^{(s)}) = b''(\omega_t^{(s)}) = d\mu_t^{(s)}/d\omega_t^{(s)}$. We focus on exponential families commonly used in non-Gaussian time-series modelling, including the Binomial, Negative Binomial, Poisson, Gamma, Normal, and Inverse Gaussian cases \bcitep{McCullagh}. All parametrization issues are discussed in Appendix~\ref{app:exp_family_components}. Similarly to the GARMA models \bcitep{benjamin2003}, we assume that the conditional mean $\mu_t^{(s)}$ is related to a linear predictor through a monotonic twice-differentiable link function $g$.


For our proposed TGARMA model, the linear predictor includes autoregressive and moving-average terms with regime-specific orders and coefficients. In particular,
\begin{equation}
	\eta_t^{(S_t)} = g\bigl(\mu_t^{(S_t)}\bigr)
	= \mathbf{x}_t^\top \boldsymbol\beta^{(S_t)}
	+ \sum_{j=1}^{p_{S_t}} \phi_j^{(S_t)}
	\bigl[g(Y_{t-j}) - \mathbf{x}_{t-j}^\top \boldsymbol\beta^{(S_{t-j})}\bigr]
	+ \sum_{j=1}^{q_{S_t}} \vartheta_j^{(S_t)}
	\bigl[g(Y_{t-j}) - \eta_{t-j}^{(S_{t-j})}\bigr],
	\label{eq:preditor_tgarma_geral}
\end{equation}
where $\boldsymbol\beta^{(s)} = (\beta_0^{(s)}, \beta_1^{(s)}, \dots, \beta_k^{(s)})^\top \in \mathbb{R}^{k+1}$ are the regression coefficients for regime $s$ (the first component being the intercept), $\boldsymbol\phi^{(s)} = (\phi_1^{(s)},\dots,\phi_{p_s}^{(s)})^\top \in \mathbb{R}^{p_s}$ are the autoregressive parameters for regime $s$, and $\boldsymbol\vartheta^{(s)} = (\vartheta_1^{(s)},\dots,\vartheta_{q_s}^{(s)})^\top \in \mathbb{R}^{q_s}$ are the moving average parameters for regime $s$. The vector of dynamic parameters is denoted by $\boldsymbol\psi =
(\boldsymbol\beta^{(1)\top}, \boldsymbol\phi^{(1)\top},
\boldsymbol\vartheta^{(1)\top}, \dots,
\boldsymbol\beta^{(R)\top}, \boldsymbol\phi^{(R)\top},
\boldsymbol\vartheta^{(R)\top})^\top
\in \Psi \subset \mathbb{R}^{\sum_{s=1}^{R}(k+1+p_s+q_s)}.$ When the regime-specific dispersion parameters are unknown, collect them in the vector $\boldsymbol\varphi =
(\varphi^{(1)},\dots,\varphi^{(R)})^\top
\in \Phi^R.$
The full continuous parameter vector for the general exponential-family formulation is $\boldsymbol\theta =
(\boldsymbol\psi^\top,\boldsymbol\varphi^\top)^\top
\in \Theta=\Psi\times\Phi^R.$ 
The parameter space enforces the mean-domain restrictions of the chosen conditional family, positivity restrictions for dispersion parameters, and the admissibility restrictions imposed on the observation-driven recursion. 
\cite{benjamin2003} comments that the ARMA root conditions for the stationarity of the GARMA model are valid for specific cases, mainly for the identity link function, and \cite{prass2022positive} discuss that this is an open problem for the GARMA models, we are assuming that these conditions are valid for each regime to avoid non-stationarity problems.

This specification defines the TGARMA model as an observation-driven threshold extension of GARMA dynamics. The next section develops the profile partial likelihood for threshold selection  and partial likelihood-based inference for the continuous parameter vector. Conditional families admitting additional shape or precision parameters can still be used in applications, but their likelihood contributions must then be written explicitly for the chosen parametrization.

\section{Partial Maximum Likelihood Inference}
\label{sec:estimation}
Let $\{(Y_t,\mathbf x_t)\}_{t=1}^{T}$ be a sample from the TGARMA model defined in \blueeqref{eq:regime_threshold_geral}, \blueeqref{eq:ef_geral} and \blueeqref{eq:preditor_tgarma_geral}. Denote by $\boldsymbol{\theta} = (\boldsymbol{\psi}^\top,\boldsymbol{\varphi}^\top)^\top \in \Theta$ the full continuous parameter vector for the general exponential-family formulation, where $\boldsymbol{\psi}$ contains the regime-specific dynamic coefficients, with $\boldsymbol{\beta}^{(s)} \in \mathbb{R}^{k+1}$, $\boldsymbol{\phi}^{(s)} \in \mathbb{R}^{p_s}$ and $\boldsymbol{\vartheta}^{(s)} \in \mathbb{R}^{q_s}$ for $s=1,\dots,R$, and $\boldsymbol{\varphi}\in\mathbb{R}_+^R$ collects the regime-specific dispersion parameters whenever they are estimated. The parameter space $\Theta$ incorporates mean-domain restrictions, positivity constraints for dispersion parameters, and the stability restrictions imposed on the observation-driven recursion. 
Inference is based on partial maximum likelihood in the sense of \bcitep{Fokianos2004}. 
Let $m=\max\{p_1,\dots,p_R,q_1,\dots,q_R,d\}$; the first $m$ observations are treated as conditioning values. Conditional on $\F_{t-1}$, the partial likelihood function for $\boldsymbol{\theta}$ is
\[
L_T(\boldsymbol{\theta};\mathbf{r},d) = \prod_{t=m+1}^{T} f\bigl(y_t \mid \F_{t-1}, S_t(\mathbf{r},d); \boldsymbol{\theta}\bigr),
\]
where $f(\cdot)$ is the conditional density given in \blueeqref{eq:ef_geral}, with $\mu_t^{(S_t)}=g^{-1}\{\eta_t^{(S_t)}(\boldsymbol{\theta};\mathbf{r},d)\}$. The canonical parameter $\omega_t^{(S_t)}$ is determined by $\mu_t^{(S_t)}=b'(\omega_t^{(S_t)})$ and coincides with $\eta_t^{(S_t)}$ only under the canonical link. The regime indicator $S_t(\mathbf{r},d)$ is defined in \blueeqref{eq:regime_threshold_geral}. For any admissible $(\mathbf r,d)$, the regime sequence is predictable because $S_t(\mathbf r,d)$ is determined by lagged information. We write the partial log-likelihood as
\begin{equation}
	\ell_T(\boldsymbol{\theta};\mathbf r,d)
	=
	\sum_{t=m+1}^{T}
	\log f\bigl(y_t\mid\F_{t-1};S_t(\mathbf r,d),\boldsymbol{\theta}\bigr)
	=
	\sum_{t=m+1}^{T}
	\ell_t(\boldsymbol{\theta};\mathbf r,d),
	\label{loglik}
\end{equation}
where the individual contribution is
\[
\ell_t(\boldsymbol{\theta};\mathbf r,d)
=
\frac{
y_t\,\omega_t^{(S_t)}(\boldsymbol{\theta};\mathbf r,d)
-b\{\omega_t^{(S_t)}(\boldsymbol{\theta};\mathbf r,d)\}}
{\varphi^{(S_t)}}
+c\bigl(y_t,\varphi^{(S_t)}\bigr).
\]
When the chosen conditional family has no free dispersion parameter, the corresponding component is omitted from $\boldsymbol{\varphi}$ and from $\boldsymbol{\theta}$.

At this point it is useful to distinguish the two components involved in estimation. The structural component is the threshold configuration $(\mathbf r,d)$, which determines the regime sequence. The regular continuous component is $\boldsymbol{\theta}$, which collects the regime-specific coefficients and, when present, the dispersion parameters. For a fixed admissible configuration $(\mathbf r,d)$, the regime sequence is treated as given and the partial likelihood is maximized only with respect to $\boldsymbol{\theta}$. The threshold configuration is selected separately, by comparing the maximized partial likelihood values over the admissible grid. Thus, all derivatives, score vectors and information matrices introduced below are computed with respect to $\boldsymbol{\theta}$, conditional on $(\mathbf r,d)$.

Formally, for a candidate threshold configuration $(\mathbf{r},d)$, the PMLE of $\boldsymbol{\theta}$ is given by
\begin{equation}
	\widehat{\boldsymbol{\theta}}(\mathbf r,d)
	=
	\arg\max_{\boldsymbol{\theta}\in\Theta}
	\ell_T(\boldsymbol{\theta};\mathbf r,d).
	\label{eq:pmle_fixed}
\end{equation}
When the threshold parameters are unknown, estimation is carried out through a profiled partial likelihood over an admissible grid. Let $\mathcal G_T\subset\mathcal R\times\mathcal D$ denote the finite set of admissible threshold configurations considered in the search, after imposing the lag and minimum-regime-size restrictions required for estimation. For each $(\mathbf r,d)\in\mathcal G_T$, define the profiled partial log-likelihood by
\[
\ell_T^{\mathrm{prof}}(\mathbf r,d)
=
\ell_T\{\widehat{\boldsymbol\theta}(\mathbf r,d);\mathbf r,d\}.
\]
The threshold configuration is selected as
\begin{equation}
	(\widehat{\mathbf r},\widehat d)
	=
	\arg\max_{(\mathbf r,d)\in\mathcal G_T}
	\ell_T^{\mathrm{prof}}(\mathbf r,d),
	\label{eq:pmle_joint}
\end{equation}
and the reported continuous-parameter estimate is $\widehat{\boldsymbol{\theta}}=\widehat{\boldsymbol{\theta}}(\widehat{\mathbf r},\widehat d)$.

Because the threshold parameters enter the likelihood through regime indicators, they are non-regular parameters. The regular likelihood theory used below is therefore conditional on the selected threshold configuration and applies only to the continuous component of the parameter vector, as in threshold time-series likelihood arguments \bcitep{chan1990likelihood,chan1993consistency,hansen2011threshold}.

\subsection{Partial score vector}
\label{subsec:scoreVector}
For a fixed admissible threshold configuration $(\mathbf r,d)$, the partial score vector is obtained by differentiating the conditional partial log-likelihood with respect to the continuous parameter vector $\boldsymbol\theta$. The regime indicators are predictable and are therefore held fixed when differentiating with respect to $\boldsymbol\theta$, including after a configuration has been selected by profiling. Define the individual and total scores by
\begin{equation}
	U_t(\boldsymbol{\theta};\mathbf r,d)
	=
	\frac{\partial\ell_t(\boldsymbol{\theta};\mathbf r,d)}
	{\partial\boldsymbol{\theta}},
	\qquad
	U_T(\boldsymbol{\theta};\mathbf r,d)
	=
	\frac{\partial\ell_T(\boldsymbol{\theta};\mathbf r,d)}
	{\partial\boldsymbol{\theta}}
	=
	\sum_{t=m+1}^{T}U_t(\boldsymbol{\theta};\mathbf r,d).
	\label{eq:score_vector_total}
\end{equation}

Let $\psi_i$ denote the $i$th element of $\boldsymbol{\psi}$, $i=1,\ldots,d_\psi$, where $d_\psi=\dim(\boldsymbol\psi)$. Since the dynamic parameters enter $\ell_t$ through the conditional mean, or equivalently through the linear predictor, the chain rule \bcitep{nelder1972} gives
\begin{equation}
	\frac{\partial\ell_T}{\partial\psi_i}
	=\sum_{t=m+1}^{T}
	\frac{\partial\ell_t}{\partial\omega_t^{(S_t)}}
	\frac{\partial\omega_t^{(S_t)}}{\partial\mu_t^{(S_t)}}
	\frac{\partial\mu_t^{(S_t)}}{\partial\eta_t^{(S_t)}}
	\frac{\partial\eta_t^{(S_t)}}{\partial\psi_i},
	\label{eq:score_chain_tgarma}
\end{equation}
From exponential-family theory \bcitep{McCullagh},
\[
\frac{\partial\ell_t}{\partial\omega_t^{(S_t)}}=\frac{Y_t-b'(\omega_t^{(S_t)})}{\varphi^{(S_t)}}=\frac{Y_t-\mu_t^{(S_t)}}{\varphi^{(S_t)}},\qquad
\frac{\partial\omega_t^{(S_t)}}{\partial\mu_t^{(S_t)}}=\frac{1}{V_t^{(S_t)}},\qquad
\frac{\partial\mu_t^{(S_t)}}{\partial\eta_t^{(S_t)}}=\frac{1}{g'(\mu_t^{(S_t)})}\equiv D_t^{(S_t)},
\]
where $V_t^{(S_t)}= b''(\omega_t^{(S_t)})=V(\mu_t^{(S_t)})$ is the variance function. Consequently,
\begin{equation}
	\frac{\partial\ell_T}{\partial\psi_i}
	=\sum_{t=m+1}^{T}
	\frac{Y_t-\mu_t^{(S_t)}}{\varphi^{(S_t)}}\,
	\frac{D_t^{(S_t)}}{V_t^{(S_t)}}\,
	\frac{\partial\eta_t^{(S_t)}}{\partial\psi_i}.
	\label{eq:score_psi_comp}
\end{equation}
For the canonical link, $\eta_t^{(S_t)}=\omega_t^{(S_t)}$, which implies $D_t^{(S_t)}/V_t^{(S_t)}=1$ and the factor simplifies to $(Y_t-\mu_t^{(S_t)})/\varphi^{(S_t)}$.

The derivatives $\partial\eta_t^{(S_t)}/\partial\psi_i$ are obtained recursively from the predictor structure in \ref{eq:preditor_tgarma_geral}, as in the GARMA framework \citep{benjamin2003}. The initial predictors and the initial derivative values are treated as fixed conditioning quantities; equivalently, all derivatives below are initialized at zero for $t\le m$. Define
\[
I_t^{(s)}=\mathbf{1}\{S_t(\mathbf{r},d)=s\},\qquad s=1,\dots,R.
\]
For each regime $s$, let $\boldsymbol{\beta}^{(s)}\in\mathbb{R}^{k+1}$, $\boldsymbol{\phi}^{(s)}\in\mathbb{R}^{p_s}$ and $\boldsymbol{\vartheta}^{(s)}\in\mathbb{R}^{q_s}$. For $t=m+1,\ldots,T$, differentiation of \ref{eq:preditor_tgarma_geral} gives
\begin{align}
	\frac{\partial\eta_t^{(S_t)}}{\partial\boldsymbol{\beta}^{(s)}}
	&= I_t^{(s)}\mathbf{x}_t
	-\sum_{j=1}^{p_{S_t}}\phi_j^{(S_t)}I_{t-j}^{(s)}\mathbf{x}_{t-j}
	-\sum_{j=1}^{q_{S_t}}\vartheta_j^{(S_t)}\frac{\partial\eta_{t-j}^{(S_{t-j})}}{\partial\boldsymbol{\beta}^{(s)}}, \label{eq:der_beta}\\[4pt]
	\frac{\partial\eta_t^{(S_t)}}{\partial\phi_l^{(s)}}
	&= I_t^{(s)}\bigl[g(Y_{t-l})-\mathbf{x}_{t-l}^{\top}\boldsymbol{\beta}^{(S_{t-l})}\bigr]
	-\sum_{j=1}^{q_{S_t}}\vartheta_j^{(S_t)}\frac{\partial\eta_{t-j}^{(S_{t-j})}}{\partial\phi_l^{(s)}},\qquad l=1,\dots,p_s,\label{eq:der_phi}\\[4pt]
	\frac{\partial\eta_t^{(S_t)}}{\partial\vartheta_l^{(s)}}
	&= I_t^{(s)}\bigl[g(Y_{t-l})-\eta_{t-l}^{(S_{t-l})}\bigr]
	-\sum_{j=1}^{q_{S_t}}\vartheta_j^{(S_t)}\frac{\partial\eta_{t-j}^{(S_{t-j})}}{\partial\vartheta_l^{(s)}},\qquad l=1,\dots,q_s.\label{eq:der_theta}
\end{align}
These recursions define the entries of the derivative matrices used below and make explicit how the observation-driven structure transmits derivative information through the lagged predictors.

Turning to the dispersion parameters, suppose that $\varphi^{(s)}$ is a free parameter in regime $s$ under the exponential-dispersion representation in \ref{eq:ef_geral}. Since $\varphi^{(s)}$ does not enter the predictor in this parametrization, direct differentiation yields
\begin{equation}
	\frac{\partial\ell_t}{\partial\varphi^{(s)}}
	=
	I_t^{(s)}\left[
	-\frac{Y_t\omega_t^{(s)}-b(\omega_t^{(s)})}{[\varphi^{(s)}]^2}
	+\frac{\partial}{\partial\varphi^{(s)}}c\bigl(Y_t,\varphi^{(s)}\bigr)
	\right].
	\label{eq:score_varphi_comp}
\end{equation}
This expression is valid for the exponential-family formulation in \blueeqref{eq:ef_geral}, in which the dispersion parameter does not enter the linear predictor. If a fitted conditional family uses additional shape or precision parameters beyond $\varphi^{(s)}$, then the corresponding score contributions must be derived from the family-specific log-likelihood rather than from \blueeqref{eq:ef_geral} alone.

For the matrix representation of the score, let $n^\ast=T-m$ denote the effective sample size. For the realized sample, define the vectors
\[
\mathbf{y}=(y_{m+1},\dots,y_T)^{\top},\qquad
\boldsymbol{\mu}=(\mu_{m+1}^{(S_{m+1})},\dots,\mu_T^{(S_T)})^{\top},
\]
and the diagonal matrices
\[
\mathcal{D}=\operatorname{diag}\bigl(D_{m+1}^{(S_{m+1})},\dots,D_T^{(S_T)}\bigr),\quad
\mathcal{V}=\operatorname{diag}\bigl(V_{m+1}^{(S_{m+1})},\dots,V_T^{(S_T)}\bigr),\quad
W=\mathcal{D}\mathcal{V}^{-1}\mathcal{D}.
\]
For each regime $s$, construct the derivative matrices
\[
B^{(s)}=\left(\frac{\partial\eta_{m+i}^{(S_{m+i})}}{\partial\beta_j^{(s)}}\right)_{\substack{1\le i\le n^{\ast}\\ 1\le j\le k+1}},\quad
P^{(s)}=\left(\frac{\partial\eta_{m+i}^{(S_{m+i})}}{\partial\phi_j^{(s)}}\right)_{\substack{1\le i\le n^{\ast}\\ 1\le j\le p_s}},\quad
H^{(s)}=\left(\frac{\partial\eta_{m+i}^{(S_{m+i})}}{\partial\vartheta_j^{(s)}}\right)_{\substack{1\le i\le n^{\ast}\\ 1\le j\le q_s}},
\]
where the entries are given by \blueeqref{eq:der_beta}--\blueeqref{eq:der_theta}. The block $H^{(s)}$ denotes the derivative matrix associated with the moving-average parameters and should not be confused with the observed Hessian $\mathcal H_T$ introduced below. The global derivative matrix for $\boldsymbol{\psi}$ is
\begin{equation}
\Delta=\bigl[ B^{(1)}\;P^{(1)}\;H^{(1)}\;\cdots\;B^{(R)}\;P^{(R)}\;H^{(R)} \bigr]\in\mathbb{R}^{n^{\ast}\times\dim(\boldsymbol{\psi})}.
\label{eq:Delta}
\end{equation}
Using \blueeqref{eq:score_psi_comp} and summing over $t$, define the diagonal matrix of inverse regime-specific dispersion parameters as
\[
\mathcal{\Phi}^{-1}
=
\operatorname{diag}
\left(
[\varphi^{(S_{m+1})}]^{-1},\ldots,[\varphi^{(S_T)}]^{-1}
\right).
\]
Then the score vector for $\boldsymbol{\psi}$ can be written compactly as
\begin{equation}
	U_{\boldsymbol{\psi}}(\boldsymbol{\theta};\mathbf r,d)=\Delta^{\top}\mathcal{\Phi}^{-1}\mathcal{D}\mathcal{V}^{-1}(\mathbf{y}-\boldsymbol{\mu}).
	\label{eq:score_matrix_psi}
\end{equation}
The compact expression in \blueeqref{eq:score_matrix_psi} is used below to derive the contribution of the dynamic parameters to the conditional information matrix in Subsection~\ref{subsec:fisher_info}.

Define the dispersion score vector by
\[
U_{\boldsymbol{\varphi}}(\boldsymbol{\theta};\mathbf r,d)=\left(U_{\varphi^{(1)}}(\boldsymbol{\theta};\mathbf r,d),\dots,U_{\varphi^{(R)}}(\boldsymbol{\theta};\mathbf r,d)\right)^\top,
\]
with $U_{\varphi^{(s)}}(\boldsymbol{\theta};\mathbf r,d)=\sum_{t=m+1}^T \partial\ell_t/\partial\varphi^{(s)}$. Finally, the full score vector is
\[
U_T(\boldsymbol{\theta};\mathbf r,d)=\bigl(U_{\boldsymbol{\psi}}(\boldsymbol{\theta};\mathbf r,d)^{\top},\;U_{\boldsymbol{\varphi}}(\boldsymbol{\theta};\mathbf r,d)^{\top}\bigr)^{\top}.
\]
For a given admissible $(\mathbf r,d)$, the PMLE $\widehat{\boldsymbol{\theta}}(\mathbf r,d)$ is obtained by maximizing the partial log-likelihood in \blueeqref{eq:pmle_fixed}. If the maximizer is interior, it satisfies
\[
U_T\{\widehat{\boldsymbol{\theta}}(\mathbf r,d);\mathbf r,d\}=\mathbf 0.
\]

\subsection{Conditional information matrix}
\label{subsec:fisher_info}

Throughout this subsection, let $(\mathbf r,d)$ denote an admissible threshold configuration from the profile grid. The regime indicator $S_t(\mathbf r,d)$ is $\F_{t-1}$-measurable and therefore is treated as part of the conditioning information when differentiating the partial log-likelihood with respect to $\boldsymbol\theta=(\boldsymbol\psi^\top,\boldsymbol\varphi^\top)^\top$.

The conditional information matrix represents the predictable information accumulated by the partial score process and constitutes the natural information object in observation-driven generalized time series models, where the likelihood is constructed from sequential conditional densities of $Y_t$ given $\F_{t-1}$. 
\[
U_t(\boldsymbol\theta;\mathbf r,d)
=
\frac{\partial\ell_t(\boldsymbol\theta;\mathbf r,d)}
{\partial\boldsymbol\theta},
\qquad
U_T(\boldsymbol\theta;\mathbf r,d)
=
\sum_{t=m+1}^{T}U_t(\boldsymbol\theta;\mathbf r,d).
\]
Under correct conditional specification, parameter-independent conditional support, and regularity conditions permitting interchange of differentiation and conditional integration, we have
\begin{equation}
\mathbb E_{\boldsymbol\theta}
\{U_t(\boldsymbol\theta;\mathbf r,d)\mid\F_{t-1}\}
=\mathbf 0,
\label{eq:score_mds}
\end{equation}
so that $\{U_t(\boldsymbol\theta;\mathbf r,d),\F_t\}$ forms a martingale-difference sequence under $\mathbb P_{\boldsymbol\theta}$, and $U_T(\boldsymbol\theta;\mathbf r,d)$ is a vector martingale. Its predictable quadratic variation is
\begin{equation}
\mathcal K_T(\boldsymbol\theta;\mathbf r,d)
=
\sum_{t=m+1}^{T}
\mathbb E_{\boldsymbol\theta}
\left[
U_t(\boldsymbol\theta;\mathbf r,d)
U_t(\boldsymbol\theta;\mathbf r,d)^\top
\mid\F_{t-1}
\right],
\label{eq:predictable_variation}
\end{equation}
which we term the conditional information matrix. Under the same regularity conditions, the conditional information identity yields the equivalent negative expected Hessian representation
\begin{equation}
\mathcal K_T(\boldsymbol\theta;\mathbf r,d)
=
\sum_{t=m+1}^{T}
\mathbb E_{\boldsymbol\theta}
\left[
-
\frac{\partial^2\ell_t(\boldsymbol\theta;\mathbf r,d)}
{\partial\boldsymbol\theta\,\partial\boldsymbol\theta^\top}
\mid\F_{t-1}
\right].
\label{eq:conditional_information}
\end{equation}

The corresponding observed negative Hessian contribution and accumulated observed Hessian are
\begin{equation}
\mathcal H_t(\boldsymbol\theta;\mathbf r,d)
=
-
\frac{\partial^2\ell_t(\boldsymbol\theta;\mathbf r,d)}
{\partial\boldsymbol\theta\,\partial\boldsymbol\theta^\top},
\qquad
\mathcal H_T(\boldsymbol\theta;\mathbf r,d)
=
\sum_{t=m+1}^{T}\mathcal H_t(\boldsymbol\theta;\mathbf r,d).
\label{eq:hessian_observed_def}
\end{equation}

The unconditional expected information and its normalized version are given by
\begin{equation}
\mathcal I_T(\boldsymbol\theta;\mathbf r,d)
=
\mathbb E_{\boldsymbol\theta}
\{\mathcal K_T(\boldsymbol\theta;\mathbf r,d)\}
=
\mathbb E_{\boldsymbol\theta}
\{\mathcal H_T(\boldsymbol\theta;\mathbf r,d)\},
\qquad
\mathcal I^{(n^\ast)}(\boldsymbol\theta;\mathbf r,d)
=
\frac{1}{n^\ast}\mathcal I_T(\boldsymbol\theta;\mathbf r,d),
\label{eq:expected_information}
\end{equation}
where $n^\ast=T-m$. In dependent data settings, $\mathcal K_T$ is generally random due to its dependence on the realized past through the regime sequence, covariates, predictors, and variance weights. The normalized expected information $\mathcal I^{(n^\ast)}$ is an expectation-level object. Under the regularity conditions of Subsection~\ref{subsec:asymptotic}, these quantities satisfy, at the true parameter $\boldsymbol\theta_0$,
\begin{equation}
\frac{1}{n^\ast}\mathcal H_T(\boldsymbol\theta_0;\mathbf r,d)
-
\mathcal I^{(n^\ast)}(\boldsymbol\theta_0;\mathbf r,d)
\xrightarrow{p}\mathbf 0,
\qquad
\frac{1}{n^\ast}\mathcal K_T(\boldsymbol\theta_0;\mathbf r,d)
-
\mathcal I^{(n^\ast)}(\boldsymbol\theta_0;\mathbf r,d)
\xrightarrow{p}\mathbf 0.
\label{eq:information_stability}
\end{equation}
When the expectation-level limit exists, we have
\begin{equation}
\mathcal I^{(n^\ast)}(\boldsymbol\theta_0;\mathbf r,d)
\longrightarrow
\mathcal I(\boldsymbol\theta_0;\mathbf r,d),
\label{eq:information_limit}
\end{equation}
where $\mathcal I(\boldsymbol\theta_0;\mathbf r,d)$ is assumed finite and positive definite. This limiting matrix appears in the martingale central limit theorem for $(n^\ast)^{-1/2}U_T(\boldsymbol\theta_0;\mathbf r,d)$ and determines the asymptotic covariance of the continuous-parameter PMLE.

We now use Lemma~\ref{lem:conditional_mean_score} to obtain the dynamic-parameter block of the conditional information matrix. For a fixed admissible threshold configuration $(\mathbf r,d)$, let $\boldsymbol\delta_t$ denote the transpose of the row of $\Delta$ associated with time $t$. Since the regime indicator and the recursive quantities are $\F_{t-1}$-measurable, combining the score representation in \blueeqref{eq:score_psi_comp} with Lemma~\ref{lem:conditional_mean_score} gives the individual dynamic score
\[
U_t^{(\psi)}(\boldsymbol\theta;\mathbf r,d)
=
\boldsymbol\delta_t
\frac{D_t^{(S_t)}}{\varphi^{(S_t)}V_t^{(S_t)}}
(Y_t-\mu_t^{(S_t)}).
\]
The same conditional score-information identities give the corresponding one-step contribution to the dynamic information block. Hence, using the matrices $\Delta$, $\mathcal{\Phi}^{-1}$, $\mathcal D$, $\mathcal V$ and $W=\mathcal D\mathcal V^{-1}\mathcal D$ defined in Subsection~\ref{subsec:scoreVector}, summing over $t=m+1,\ldots,T$ yields

\begin{equation}
\mathcal K_T^{(\psi\psi)}(\boldsymbol\theta;\mathbf r,d)
=
\Delta^\top\mathcal{\Phi}^{-1}W\Delta.
\label{eq:fisher_psi}
\end{equation}
For the canonical link, the simplification noted after \blueeqref{eq:score_psi_comp} gives $W=\operatorname{diag}\{V_t^{(S_t)}\}_{t=m+1}^{T}$.

Equivalently, expanding $\Delta$ into its regime-specific blocks gives the partitioned conditional information matrix
\begin{equation}
\mathcal K_T^{(\psi\psi)}(\boldsymbol\theta;\mathbf r,d)
=
\begin{pmatrix}
\mathcal K_T^{(1,1)} & \mathcal K_T^{(1,2)} & \cdots & \mathcal K_T^{(1,R)} \\
\mathcal K_T^{(2,1)} & \mathcal K_T^{(2,2)} & \cdots & \mathcal K_T^{(2,R)} \\
\vdots & \vdots & \ddots & \vdots \\
\mathcal K_T^{(R,1)} & \mathcal K_T^{(R,2)} & \cdots & \mathcal K_T^{(R,R)}
\end{pmatrix},
\label{eq:fisher_psi_partitioned}
\end{equation}
The corresponding expected information blocks are obtained by applying $\mathbb E_{\boldsymbol\theta}$ to the blocks of $\mathcal K_T^{(\psi\psi)}$.

The threshold structure enters through the block decomposition of $\Delta$ in \blueeqref{eq:Delta}. Suppressing the common arguments $(\boldsymbol\theta;\mathbf r,d)$, the finite-sample conditional blocks of \blueeqref{eq:fisher_psi} are
\begin{align}
\mathcal K_{T,\beta^{(s)}\beta^{(u)}} &= B^{(s)\top}\mathcal{\Phi}^{-1}WB^{(u)}, &
\mathcal K_{T,\beta^{(s)}\phi^{(u)}} &= B^{(s)\top}\mathcal{\Phi}^{-1}WP^{(u)}, &
\mathcal K_{T,\beta^{(s)}\vartheta^{(u)}} &= B^{(s)\top}\mathcal{\Phi}^{-1}WH^{(u)}, \nonumber\\[4pt]
\mathcal K_{T,\phi^{(s)}\beta^{(u)}} &= P^{(s)\top}\mathcal{\Phi}^{-1}WB^{(u)}, &
\mathcal K_{T,\phi^{(s)}\phi^{(u)}} &= P^{(s)\top}\mathcal{\Phi}^{-1}WP^{(u)}, &
\mathcal K_{T,\phi^{(s)}\vartheta^{(u)}} &= P^{(s)\top}\mathcal{\Phi}^{-1}WH^{(u)}, \nonumber\\[4pt]
\mathcal K_{T,\vartheta^{(s)}\beta^{(u)}} &= H^{(s)\top}\mathcal{\Phi}^{-1}WB^{(u)}, &
\mathcal K_{T,\vartheta^{(s)}\phi^{(u)}} &= H^{(s)\top}\mathcal{\Phi}^{-1}WP^{(u)}, &
\mathcal K_{T,\vartheta^{(s)}\vartheta^{(u)}} &= H^{(s)\top}\mathcal{\Phi}^{-1}WH^{(u)}.
\label{eq:fisher_blocks}
\end{align}
For example, the corresponding expected $\beta^{(s)}$--$\beta^{(u)}$ component is
$\mathbb E_{\boldsymbol\theta}\{B^{(s)\top}\mathcal{\Phi}^{-1}WB^{(u)}\}$, and the remaining expected components are obtained analogously from \blueeqref{eq:fisher_blocks}.
Unlike a regime-separated GLM with independent observations, the off-diagonal blocks in \blueeqref{eq:fisher_blocks} need not vanish. Although the coefficients are regime-specific, the recursive derivatives of $\eta_t$ may depend on past predictors and residuals generated under other regimes, so information may be transmitted across regime histories. Cross-regime blocks should therefore be set to zero only under an explicitly verified orthogonality or separation restriction.

 For the dispersion component, define the individual score contribution  $a_t(\boldsymbol\theta;\mathbf r,d) = \partial\ell_t(\boldsymbol\theta;\mathbf r,d)/{\partial\boldsymbol\varphi}$. The corresponding conditional information blocks are then given by
\begin{align}
\mathcal K_T^{(\varphi\varphi)}(\boldsymbol\theta;\mathbf r,d)
&=
\sum_{t=m+1}^{T}
\mathbb E_{\boldsymbol\theta}
\left[
a_ta_t^\top
\mid\F_{t-1}
\right],\\
\mathcal K_T^{(\psi\varphi)}(\boldsymbol\theta;\mathbf r,d)
&=
\sum_{t=m+1}^{T}
\mathbb E_{\boldsymbol\theta}
\left[
U_t^{(\psi)}a_t^\top
\mid\F_{t-1}
\right].
\end{align}
For the exponential-dispersion parametrization in \blueeqref{eq:ef_geral}, $\boldsymbol\varphi$ does not enter $\eta_t$ or $\mu_t$; hence $\partial\mu_t/\partial\boldsymbol\varphi=\mathbf 0$. Under the same regularity conditions used above,
\[
\mathbb E_{\boldsymbol\theta}
\{a_t(\boldsymbol\theta;\mathbf r,d)\mid\F_{t-1}\}
=\mathbf 0,
\qquad
\mathbb E_{\boldsymbol\theta}
\{(Y_t-\mu_t)a_t(\boldsymbol\theta;\mathbf r,d)^\top
\mid\F_{t-1}\}
=\mathbf 0.
\]
The second equality follows by differentiating the conditional mean identity $\mathbb E_{\boldsymbol\theta}(Y_t\mid\F_{t-1})=\mu_t$ with respect to $\boldsymbol\varphi$, while holding $\mu_t$ fixed. Combining this identity with the dynamic score expression displayed above gives
\[
\mathcal K_T^{(\psi\varphi)}(\boldsymbol\theta;\mathbf r,d)=\mathbf 0.
\]
Thus, for the exponential-dispersion parametrization stated in \blueeqref{eq:ef_geral}, the mean/dynamic parameters and the dispersion parameters are orthogonal in the information sense of \bcitep{cox1987parameter}. Consequently, in this formulation,
\[
\mathcal K_T(\boldsymbol\theta;\mathbf r,d)
=
\begin{pmatrix}
\mathcal K_T^{(\psi\psi)}(\boldsymbol\theta;\mathbf r,d) & \mathbf 0\\
\mathbf 0 & \mathcal K_T^{(\varphi\varphi)}(\boldsymbol\theta;\mathbf r,d)
\end{pmatrix}.
\]
If the selected conditional family is parametrized with additional shape or precision parameters that affect the conditional mean, the variance function, or the recursion, this block orthogonality no longer follows automatically from \blueeqref{eq:ef_geral}. In that case the corresponding score and information blocks must be derived from the selected family-specific log-likelihood.

\subsection{Observed Hessian, OPG and covariance estimation}
\label{subsec:observed_opg}

The conditional information matrix in Subsection~\ref{subsec:fisher_info} serves as the population object for asymptotic theory. Its empirical counterparts are computed from the fitted partial likelihood. The observed negative Hessian $\mathcal H_T(\boldsymbol\theta;\mathbf r,d)$ is defined in \blueeqref{eq:hessian_observed_def}, and the outer product of individual scores is
\begin{equation}
\mathcal J_T(\boldsymbol\theta;\mathbf r,d)
=
\sum_{t=m+1}^{T}
U_t(\boldsymbol\theta;\mathbf r,d)
U_t(\boldsymbol\theta;\mathbf r,d)^\top,
\label{eq:opg_matrix}
\end{equation}
which we refer to as the outer product of gradients (OPG) matrix. Here $\mathcal J_T$ represents the optional quadratic variation associated with the score martingale. For a given admissible $(\mathbf r,d)$ and correct conditional specification,
\[
\frac{1}{n^\ast}\mathcal H_T(\boldsymbol\theta_0;\mathbf r,d)
\xrightarrow{p}\mathcal I(\boldsymbol\theta_0;\mathbf r,d),
\qquad
\frac{1}{n^\ast}\mathcal J_T(\boldsymbol\theta_0;\mathbf r,d)
\xrightarrow{p}\mathcal I(\boldsymbol\theta_0;\mathbf r,d),
\]
provided the Hessian and OPG satisfy the same uniform convergence conditions employed in the asymptotic argument below. By consistency, the same limits hold when $\boldsymbol\theta_0$ is replaced by an interior PMLE $\widehat{\boldsymbol\theta}$.

It is useful to distinguish the covariance of the unscaled estimator from the asymptotic covariance of the scaled estimator. Let
\[
\widehat{\mathcal H}_T=(n^\ast)^{-1}\mathcal H_T(\widehat{\boldsymbol\theta};\mathbf r,d),
\qquad
\widehat{\mathcal J}_T=(n^\ast)^{-1}\mathcal J_T(\widehat{\boldsymbol\theta};\mathbf r,d).
\]
Under correct conditional specification, $\widehat{\mathcal H}_T^{-1}$ estimates the covariance matrix of $\sqrt{n^\ast}\{\widehat{\boldsymbol\theta}(\mathbf r,d)-\boldsymbol\theta_0\}$, whereas $\mathcal H_T(\widehat{\boldsymbol\theta};\mathbf r,d)^{-1}$ estimates the covariance matrix of the unscaled PMLE. If the conditional density is used as a working likelihood and the information identity is not imposed, the sandwich covariance estimator for the unscaled PMLE is
\begin{equation}
\widehat{\Sigma}_T
=
\mathcal H_T(\widehat{\boldsymbol\theta};\mathbf r,d)^{-1}
\mathcal J_T(\widehat{\boldsymbol\theta};\mathbf r,d)
\mathcal H_T(\widehat{\boldsymbol\theta};\mathbf r,d)^{-1}.
\label{eq:sandwich}
\end{equation}
Equivalently, $n^\ast\widehat{\Sigma}_T=\widehat{\mathcal H}_T^{-1}\widehat{\mathcal J}_T\widehat{\mathcal H}_T^{-1}$ estimates the covariance of the scaled estimator. All covariance matrices in this subsection are conditional on the threshold configuration used in the fit and do not incorporate the non-regular uncertainty arising from selecting $(\mathbf r,d)$.

\subsection{Asymptotic properties}
\label{subsec:asymptotic}

 The large-sample result below concerns the regular continuous component of the PMLE, conditional on an admissible threshold configuration $(\mathbf r,d)$. The argument combines martingale central limit theory \bcitep{hall1980martingale} with the partial-likelihood framework for observation-driven generalized time series \bcitep{Fokianos2004}. The threshold-specific admissibility and identifiability conditions are motivated by the threshold time-series literature \bcitep{chan1990likelihood,chan1993consistency,hansen2011threshold}. The assumptions are written explicitly because the threshold recursion makes stability and identifiability model-dependent. The conditions below are sufficient for the regular PMLE argument; they are not intended to be minimal.

\begin{enumerate}
\item[(A1)] The threshold configuration $(\mathbf r,d)$ is fixed and admissible, with $d\ge 1$, and $S_t(\mathbf r,d)$ is $\F_{t-1}$-measurable. Following the
  usual use of admissible or trimmed threshold regions in threshold estimation \bcitep{hansen2000sample,hansen2011threshold}, each active regime is required to contain a non-
  vanishing fraction of the effective sample. Specifically, if $n_s(\mathbf r,d)=\sum_{t=m+1}^{T}I_t^{(s)}(\mathbf r,d)$, then $n_s(\mathbf r,d)/n^\ast$ is bounded away from zero in probability for each active regime $s$.
\item[(A2)] The true continuous parameter
$\boldsymbol\theta_0$ belongs to the interior of the parameter space
$\Theta$, and the parameter space enforces the support, positivity,
mean-domain and admissibility restrictions required by the selected
conditional family and recursion.
\item[(A3)] The conditional distribution is correctly specified and has
parameter-independent support. In a neighbourhood of
$\boldsymbol\theta_0$, the conditional log-density is twice continuously
differentiable in $\boldsymbol\theta$, with first and second derivatives
dominated by conditionally integrable random variables. The interchanges of
differentiation and conditional integration used in
\blueeqref{eq:score_mds}--\blueeqref{eq:predictable_variation} are valid.
\item[(A4)] For the admissible configuration $(\mathbf r,d)$, the TGARMA recursion admits a well-defined stationary and ergodic solution at $\boldsymbol\theta_0$. The required moments of the score and Hessian contributions are finite, and the initial conditioning values have asymptotically negligible effect on $(n^\ast)^{-1}\ell_T$ and $(n^\ast)^{-1/2}U_T$.
\item[(A5)] The model is identifiable for the admissible configuration
$(\mathbf r,d)$: if
$\ell_t(\boldsymbol\theta;\mathbf r,d)$ and
$\ell_t(\boldsymbol\theta_0;\mathbf r,d)$ induce the same conditional
distribution almost surely for all large $t$, then
$\boldsymbol\theta=\boldsymbol\theta_0$.
\item[(A6)] A uniform law of large numbers holds for the normalized
partial log-likelihood and for the normalized observed Hessian in a
neighbourhood of $\boldsymbol\theta_0$. In particular, for any sequence
$\bar{\boldsymbol\theta}\xrightarrow{p}\boldsymbol\theta_0$,
$(n^\ast)^{-1}\mathcal H_T(\bar{\boldsymbol\theta};\mathbf r,d)
\xrightarrow{p}\mathcal I(\boldsymbol\theta_0;\mathbf r,d)$.
\item[(A7)] The martingale-difference array
$U_t(\boldsymbol\theta_0;\mathbf r,d)$ satisfies the conditional Lindeberg
condition, and its predictable variation satisfies
\[
\frac{1}{n^\ast}\mathcal K_T(\boldsymbol\theta_0;\mathbf r,d)
\xrightarrow{p}
\mathcal I(\boldsymbol\theta_0;\mathbf r,d),
\]
where $\mathcal I(\boldsymbol\theta_0;\mathbf r,d)$ is finite and positive
definite.
\end{enumerate}

\begin{theorem}[Conditional asymptotic normality of the continuous-parameter PMLE]
\label{thm:continuous_pmle}
Under (A1)--(A7), the probability that a locally unique interior PMLE
$\widehat{\boldsymbol\theta}(\mathbf r,d)$ exists in a neighbourhood of
$\boldsymbol\theta_0$ tends to one. Any such sequence of local maximizers is
consistent,
\[
\widehat{\boldsymbol\theta}(\mathbf r,d)
\xrightarrow{p}
\boldsymbol\theta_0,
\]
and asymptotically normal:
\begin{equation}
\sqrt{n^\ast}
\{\widehat{\boldsymbol\theta}(\mathbf r,d)-\boldsymbol\theta_0\}
\xrightarrow{d}
\mathcal N\left(
\mathbf 0,
\mathcal I(\boldsymbol\theta_0;\mathbf r,d)^{-1}
\right).
\label{eq:asymptotic_normality}
\end{equation}
\end{theorem}

The proof is given in Appendix~\ref{app:continuous_pmle_proof}. The result is conditional on the threshold configuration used in the fit and does not provide a limiting distribution for $(\widehat{\mathbf r},\widehat d)$. Inference on the threshold configuration itself requires the corresponding non-regular profile-likelihood or resampling theory.

\section{Confidence Intervals and Hypothesis Testing}
\label{sec:asymptotics}

The asymptotic normality result in Subsection~\ref{subsec:asymptotic} provides the basis for confidence intervals and Wald tests for the continuous parameters of the TGARMA model. For a fixed admissible threshold configuration $(\mathbf r,d)$, let $\widehat{\Sigma}_T(\mathbf r,d)$ denote an estimator of the unscaled covariance matrix of $\widehat{\boldsymbol\theta}(\mathbf r,d)$. Under correct conditional specification,
\[
\widehat{\Sigma}_T(\mathbf r,d)
=
\mathcal H_T\{\widehat{\boldsymbol\theta}(\mathbf r,d);\mathbf r,d\}^{-1},
\]
or, equivalently,
$\widehat{\Sigma}_T(\mathbf r,d)=(n^\ast)^{-1}\widehat{\mathcal I}_T(\mathbf r,d)^{-1}$, where
$\widehat{\mathcal I}_T(\mathbf r,d)=(n^\ast)^{-1}\mathcal H_T\{\widehat{\boldsymbol\theta}(\mathbf r,d);\mathbf r,d\}$. If the conditional distribution is used only as a working model, the sandwich estimator in \blueeqref{eq:sandwich} is used instead. To simplify notation, the dependence on $(\mathbf r,d)$ is suppressed below.

For the $i$th continuous parameter, the Wald pivot satisfies
\[
\frac{\widehat{\theta}_i-\theta_{0i}}
{\sqrt{(\widehat{\Sigma}_T)_{ii}}}
\xrightarrow{d}
\mathcal N(0,1),
\qquad i=1,\ldots,d_\theta,
\]
where $d_\theta=\dim(\boldsymbol\theta)$. Hence an approximate
$100(1-\alpha)\%$ confidence interval for $\theta_i$ is
\[
\left[
\widehat{\theta}_i
-z_{1-\alpha/2}\sqrt{(\widehat{\Sigma}_T)_{ii}},
\;
\widehat{\theta}_i
+z_{1-\alpha/2}\sqrt{(\widehat{\Sigma}_T)_{ii}}
\right],
\]
where $z_{1-\alpha/2}$ denotes the $(1-\alpha/2)$ quantile of the standard normal distribution.

For the scalar null hypothesis $\mathcal H_0:\theta_i=\theta_i^0$, the Wald statistic
\[
z_i
=
\frac{\widehat{\theta}_i-\theta_i^0}
{\sqrt{(\widehat{\Sigma}_T)_{ii}}}
\]
is asymptotically standard normal under $\mathcal H_0$ \bcitep{wald1943tests,pawitan}. More generally, for $q$ linear restrictions $R\boldsymbol\theta=\mathbf c$, with $R$ of full row rank $q$,
\[
W_T
=
(R\widehat{\boldsymbol\theta}-\mathbf c)^\top
\{R\widehat{\Sigma}_T R^\top\}^{-1}
(R\widehat{\boldsymbol\theta}-\mathbf c)
\xrightarrow{d}\chi_q^2 .
\]
These inferential statements apply only to the continuous parameter vector, conditionally on the threshold configuration used in the fit. They should not be used for $(\mathbf r,d)$, nor for testing the existence of a threshold effect, because such hypotheses involve discontinuity or non-identification under the null. Threshold inference requires separate non-regular profile-likelihood, bootstrap or subsampling arguments \bcitep{chan1993consistency,hansen2000sample,hansen2011threshold}.

\section{Properties of the TGARMA model}
\label{sec:properties}

This section summarizes structural features that are useful for interpreting and implementing the TGARMA class. For a fixed admissible threshold configuration $(\mathbf r,d)$, the recursion is observation-driven: once the past and the covariate path are fixed, the active regime and the conditional distribution of $Y_t$ are determined up to the continuous parameter vector. The statements below are deliberately limited. The first establishes finite-sample well-definedness of the conditional recursion, whereas stability and marginal moments are discussed as model-dependent properties. They should not be read as a complete stationarity theory for every distributional, link-function and threshold specification in the class.

\begin{proposition}[Well-defined conditional recursion]
\label{prop:well_defined_recursion}
Fix an admissible threshold configuration $(\mathbf r,d)$. Suppose that the covariate path is finite and predictable, or deterministic, the initial predictors are finite, the transformed lagged observations $g(Y_{t-j})$ entering \blueeqref{eq:preditor_tgarma_geral} are finite almost surely under the adopted boundary convention, and every generated predictor is mapped by the inverse link into the mean domain of the chosen conditional family. Then, conditional on the first $m$ observations and on the covariate path, \blueeqref{eq:preditor_tgarma_geral} defines a unique finite sequence of predictors $\eta_t^{(S_t)}$ and conditional means $\mu_t^{(S_t)}=g^{-1}\{\eta_t^{(S_t)}\}$ for $t=m+1,\ldots,T$.
\end{proposition}

\noindent\textit{Proof.}
For a fixed admissible $(\mathbf r,d)$, the regime indicator $S_t(\mathbf r,d)$ is $\F_{t-1}$-measurable. Given the initial values and the covariate path, the predictor at each $t>m$ is a finite deterministic function of predictable covariates, finitely many transformed lagged observations and finitely many previously constructed predictors. The asserted sequence is therefore obtained recursively and is unique by induction from the finite initial predictors. This finite-sample construction does not, by itself, imply stationarity or ergodicity of the infinite process. \qed

\subsection{Stability conditions}
Let
\[
A_s=\sum_{j=1}^{p_s}|\phi_j^{(s)}|,
\qquad
M_s=\sum_{j=1}^{q_s}|\vartheta_j^{(s)}|,
\qquad
\rho=\max_{1\le s\le R}(A_s+M_s).
\]

The condition
\begin{equation}
\rho<1
\label{eq:rho_condition}
\end{equation}
is a convenient uniform contraction restriction for the feedback coefficients in the linear predictor. Because the bound is taken over all regimes, it controls the deterministic propagation of lagged predictor and transformed-response terms along every admissible realized regime path. If the covariates are bounded and the transformed lagged observations have the required moments, \blueeqref{eq:rho_condition} helps rule out explosive accumulation in the linear feedback component; full stochastic stability still depends on the complete observation-driven recurrence.

Condition \blueeqref{eq:rho_condition} should therefore be interpreted as an admissibility and stability diagnostic, not as a necessary and sufficient stationarity theorem for the full TGARMA class. For non-Gaussian observation-driven threshold models, existence, uniqueness of a stationary solution and ergodicity depend on the full stochastic recurrence, including the link, the conditional distribution, the transition variable, the covariate process and any boundary convention required to evaluate $g(Y_t)$. In this paper, stability, ergodicity, finite moments and nonsingularity of the limiting information matrix are imposed as regularity conditions for PMLE, following the observation-driven likelihood framework of \bcitep{Fokianos2004} and the threshold time-series literature \bcitep{tong1990,chan1993consistency,hansen2011threshold}.

The link must also respect the support of the chosen conditional family. For nonnegative responses under the log link, $\mu_t^{(S_t)}=\exp\{\eta_t^{(S_t)}\}>0$ automatically, but transformed lagged observations such as $g(Y_{t-j})$ still require a well-defined boundary convention whenever the support includes boundary values. Under the identity link, positivity of the predictor must be imposed directly for all admissible histories and covariates whenever the conditional family has a positive mean domain; otherwise the likelihood is not well defined.

\subsection{Moment statements}
Closed-form marginal moments are generally unavailable for nonlinear threshold GARMA models. Let $I_t^{(s)}=\mathds{1}\{S_t(\mathbf r,d)=s\}$. Since the regime indicator is $\F_{t-1}$-measurable, the conditional mean satisfies $\mathbb E(Y_t\mid\F_{t-1})=\sum_{s=1}^{R}I_t^{(s)}\mu_t^{(s)}$. Hence, by the tower property,
\[
\mathbb E(Y_t)
=
\sum_{s=1}^{R}
\mathbb E\!\left[
I_t^{(s)}\mu_t^{(s)}
\right],
\]
whenever the expectation exists. This identity does not assume independence between $I_t^{(s)}$ and $\mu_t^{(s)}$. In general it cannot be simplified to a weighted average of regime intercepts, because both quantities are jointly determined by the lagged process and by the observation-driven recursion. When the link is non-identity, even the marginal mean depends on the stationary distribution of the predictor.

A weighted-intercept expression is available only under additional restrictions. Suppose the process is stationary, the required expectations are finite, the covariates are constant with $\mathbf x_t^\top\boldsymbol\beta^{(s)}=\beta_0^{(s)}$, the link is the identity, and the regime-weighted dynamic contributions in \blueeqref{eq:preditor_tgarma_geral} have zero marginal expectation under the stationary law. If $\pi_s=\mathbb P(S_t=s)$, then
\[
\mathbb{E}(Y_t)=\sum_{s=1}^{R}\pi_s\beta_0^{(s)}.
\]
This expression is a restricted benchmark, not a general moment formula for TGARMA models. For log-link, nonlinear-link or count-data specifications, marginal moments should be evaluated through model-specific analytical arguments, simulation from the fitted recursion, or both \bcitep{benjamin2003,Fokianos2004}.

\section{Model Selection, Diagnostics and Forecasting}
\label{sec:implementation}

\subsection{Model selection}
\label{subsec:model_selection}

Model selection is carried out over a finite admissible class $\mathfrak M$ of TGARMA specifications. Each candidate
\[
\mathcal M=(p_1,\ldots,p_R,q_1,\ldots,q_R,d,\mathbf r)\in\mathfrak M
\]
determines the regime-specific GARMA orders and the threshold configuration. For each $\mathcal M$, the continuous parameter vector is estimated by PMLE, conditionally on the threshold configuration implied by that candidate.

For each candidate specification, the partial likelihood is optimized numerically using BFGS-type quasi-Newton optimization \bcitep{press1992numerical}, with analytical first derivatives when available. Following standard GARMA practice \bcitep{benjamin2003,Bayers,melo2020,sales2022berg}, starting values are obtained by setting the moving-average parameters to zero and fitting, within each regime, an initial generalized linear model with response $Y_t$ and linear predictor containing $\mathbf{x}_t$ and the available transformed lagged responses $g(Y_{t-1}),\ldots,g(Y_{t-p_s})$. The resulting estimates initialize $\boldsymbol{\beta}^{(s)}$ and $\boldsymbol{\phi}^{(s)}$, while $\boldsymbol{\vartheta}^{(s)}$ is initialized at zero. Dispersion parameters are initialized at admissible values for the selected conditional family.

To compare specifications with different lag orders on the same likelihood support, let $m_{\mathrm{sel}}$ denote the largest lag required by any candidate in $\mathfrak M$, and set $n_{\mathrm{sel}}=T-m_{\mathrm{sel}}$. The selection log-likelihood is evaluated on the common window $t=m_{\mathrm{sel}}+1,\ldots,T$. Denote by
\[
\widehat{\boldsymbol\theta}_{\mathrm{sel}}(\mathcal M)
=
\arg\max_{\boldsymbol\theta_{\mathcal M}\in\Theta_{\mathcal M}}
\ell_{\mathrm{sel}}(\boldsymbol\theta_{\mathcal M};\mathcal M)
\]
the PMLE of the continuous parameter vector under candidate $\mathcal M$, where $\ell_{\mathrm{sel}}$ is the partial log-likelihood restricted to the common selection window. The corresponding profiled partial log-likelihood is
\[
\ell_{\mathrm{sel}}^{\mathrm{prof}}(\mathcal M)
=
\ell_{\mathrm{sel}}\{
\widehat{\boldsymbol\theta}_{\mathrm{sel}}(\mathcal M);\mathcal M
\}.
\]

The profile information criteria used for comparison are
\[
\mathrm{AIC}_{\mathrm{sel}}(\mathcal M)
=
-2\ell_{\mathrm{sel}}^{\mathrm{prof}}(\mathcal M)
+2d_\theta(\mathcal M),
\]
and
\[
\mathrm{BIC}_{\mathrm{sel}}(\mathcal M)
=
-2\ell_{\mathrm{sel}}^{\mathrm{prof}}(\mathcal M)
+d_\theta(\mathcal M)\log n_{\mathrm{sel}},
\]
where $d_\theta(\mathcal M)=\sum_{s=1}^{R}(k+1+p_s+q_s)+d_{\varphi}(\mathcal M)$ denotes the number of continuous parameters estimated by PMLE under $\mathcal M$, and $d_{\varphi}(\mathcal M)$ is the number of estimated regime-specific dispersion parameters. The threshold vector $\mathbf r$ and the delay $d$ are not included in $d_\theta(\mathcal M)$ because they are selected over the admissible grid and enter the likelihood through discontinuous regime indicators. The selected specification is the admissible candidate minimizing the chosen criterion, for example
\[
\widehat{\mathcal M}_{\mathrm{BIC}}
=
\arg\min_{\mathcal M\in\mathfrak M}
\mathrm{BIC}_{\mathrm{sel}}(\mathcal M),
\qquad
\widehat{\mathcal M}_{\mathrm{AIC}}
=
\arg\min_{\mathcal M\in\mathfrak M}
\mathrm{AIC}_{\mathrm{sel}}(\mathcal M).
\]

The admissible class $\mathfrak M$ is restricted so that each regime contains enough effective observations, within the common selection window, to identify the corresponding regime-specific continuous parameters. The criteria above are used as finite-sample profile-likelihood selection devices; they do not provide regular Wald-type inference for $(\mathbf r,d)$. After model selection, standard errors and Wald-type intervals are computed only for the continuous parameter vector, conditionally on the selected specification.

 For the selected model, the maximized partial log-likelihood, the selected threshold configuration, the effective regime sizes, the stability diagnostic $\widehat\rho$, and
  the conditioning of the observed information matrix should be inspected together with the residual diagnostics below.

\subsection{Residual analysis}
\label{subsec:residual_analysis}
Residual diagnostics assess both the conditional mean and the conditional distribution \bcitep{Fokianos2004}. Let $\widehat F_t(y)=
F\{y\mid\F_{t-1},S_t(\widehat{\mathbf r},\widehat d);
\widehat{\boldsymbol\theta}(\widehat{\mathbf r},\widehat d)\}$
be the fitted one-step conditional distribution. The raw residual $e_t=y_t-\widehat{\mu}_t$ is used to assess the martingale-difference implication of the conditional-mean specification. Since the fitted conditional variance is generally time-varying, define
\[
\widehat{\sigma}_t^2
=
\operatorname{Var}_{\widehat{\boldsymbol\theta}}
\{Y_t\mid\F_{t-1},S_t=\widehat s_t\},
\qquad
\widehat s_t=S_t(\widehat{\mathbf r},\widehat d),
\]
and use the Pearson residual
\[
r_t^{(P)}
=
\frac{y_t-\widehat\mu_t}{\widehat\sigma_t}.
\]
 The latter is the primary scale-adjusted residual for checking remaining conditional-mean misspecification, omitted covariates and regime-specific lack of fit. The formal implication of the model is that $Y_t-\mu_t$ is a martingale-difference sequence with respect to the  the
filtration $\F_t$ defined in Subsection~\ref{subsec:proposed_model}. Therefore, martingale-difference tests, such as the Dominguez--Lobato test, are natural diagnostics for this implication.For true errors with finite second moments, the martingale-difference property implies zero autocovariances at every positive lag, regardless of the link. Residual ACF and PACF plots remain useful, but the reference distribution of Ljung--Box statistics on fitted residuals requires justification for estimation and selection effects \bcitep{ljung1978measure}.

Distributional checks use probability integral transforms. For continuous conditional distributions,
\[
r_t^{(q)}=\Phi_N^{-1}\{\widehat F_t(y_t)\},
\]
where $\Phi_N^{-1}$ is the standard normal quantile function. For discrete conditional distributions, let $\widehat F_t(y_t^-)$ be the left limit of the fitted conditional distribution at $y_t$; for integer-valued data this is $\widehat F_t(y_t-1)$. The randomized quantile residual is
\[
r_t^{(rq)}=\Phi_N^{-1}(u_t),
\qquad
u_t\mid y_t\sim
\operatorname{Uniform}\{\widehat F_t(y_t^-),\widehat F_t(y_t)\}.
\]
Under correct conditional specification and regular estimation, these residuals should be approximately standard normal and should not display systematic serial dependence. QQ-plots, PIT histograms, residual ACF plots and regime-stratified residual summaries are therefore natural complements to formal tests \bcitep{dunn1996randomized,li2003diagnostic,koul2005model}.

\subsection{Forecasting}
\label{subsec:forecasting}
 Forecasts are obtained by recursively evaluating the fitted conditional mean and, for predictive intervals, the fitted conditional distribution. Suppose the sample ends at $T$ and forecasts are required for $h=1,\ldots,h_0$. Let $\widehat S_{T+h}$ denote the regime used in the forecasting recursion. If the threshold variable determining $S_{T+h}$ is already observed at time $T$, then $\widehat S_{T+h}=S_{T+h}$. If it depends on future unobserved quantities, the regime must be generated by either a plug-in path or a simulated predictive path.

 approximate plug-in forecasting, define $\widetilde y_t=y_t$ for $t\le T$. After computing the forecast at horizon $h$, set $\widetilde y_{T+h}=\widehat\mu_{T+h}$ before moving to the next horizon. The plug-in recursion is
\[
\widehat{\eta}_{T+h}^{(\widehat S_{T+h})}
=\mathbf{x}_{T+h}^\top\widehat{\boldsymbol\beta}^{(\widehat S_{T+h})}
+\sum_{i=1}^{p_{\widehat S_{T+h}}}
\widehat\phi_i^{(\widehat S_{T+h})}
\left[
g(\widetilde y_{T+h-i})
-\mathbf{x}_{T+h-i}^\top
\widehat{\boldsymbol\beta}^{(\widehat S_{T+h-i})}
\right]
+\sum_{j=1}^{q_{\widehat S_{T+h}}}
\widehat\vartheta_j^{(\widehat S_{T+h})}
\left[
g(\widetilde y_{T+h-j})
-\widehat\eta_{T+h-j}^{(\widehat S_{T+h-j})}
\right],
\]
with
\[
\widehat\mu_{T+h}=g^{-1}\{\widehat\eta_{T+h}^{(\widehat S_{T+h})}\}.
\]
In simulation-based forecasting, future responses are instead generated from the fitted conditional distribution and all transformed lags and regimes are updated along each simulated path. When future regimes depend on future responses, plug-in forecasts ignore part of the predictive uncertainty. Predictive intervals should therefore be obtained by parametric simulation: simulate future paths, update regimes using the same threshold rule, and compute empirical quantiles of the simulated $Y_{T+h}$ values. Parameter uncertainty can be added by drawing $\boldsymbol\theta$ from the asymptotic normal approximation based on the covariance estimator in Subsection~\ref{subsec:observed_opg}, conditionally on the selected threshold configuration and subject to the admissibility restrictions of the fitted model. Forecast accuracy should be evaluated on a common forecast origin using RMSE, MAE, MAPE.

\section{Simulation study}
\label{sec:simulation}
 
We conduct a Monte Carlo study to assess the finite-sample performance of the PMLE for the continuous parameters of two-regime negative binomial TGARMA models. We adopted settings similar to the application study (see Section~\ref{sec:application} ). Two intercept-only specifications are considered: TGAR(1), with $p_1=p_2=1$ and $q_1=q_2=0$, and TGARMA(1,1), with $p_1=p_2=q_1=q_2=1$. Both use the logarithmic link, with common generating values $\beta_0^{(1)}=\beta_0^{(2)}=7,\qquad
(\phi_1^{(1)},\phi_1^{(2)})=(0.20,0.50), (\kappa_1,\kappa_2)=(179,427).$
For TGARMA(1,1), $(\theta_1^{(1)},\theta_1^{(2)})=(0.15,0.20)$. The active regime is determined by $Y_{t-2}\leq 1135$ or $Y_{t-2}>1135$. Orders, threshold and delay are fixed at their generating values during estimation. For each specification and sample size $n\in\{200,500,1000\}$, we perform 5,000 replications after discarding 500 burn-in observations. Generation and estimation use $\log\{\max(Y_t,0.3)\}$ for lagged counts. The first two retained observations are conditioned upon, yielding $n-2$ likelihood contributions. Estimation is carried out in R using BFGS with analytical first derivatives. Within-regime negative binomial GLMs based on centered log-transformed lagged counts supply starting values for the intercepts, autoregressive coefficients and shape parameters; moving-average coefficients, when present, start at zero. Nominal 95\% Wald intervals use sandwich covariance estimates, with shape-parameter intervals constructed on the log scale and then exponentiated. Table~\ref{tab:tgar_tgarma_simulation_glm} reports the mean, percentage relative bias (RB), empirical variance (Var), mean squared error (MSE) and coverage rate (CR).
Empirical variances and MSEs decrease with sample size for every parameter in both specifications. Intercept estimates remain close to their generating values, whereas shape estimates exhibit upward bias, particularly in the second regime at $n=200$. For $\kappa_2$, RB decreases from 12.55\% to 2.06\% in TGAR(1), and from 14.79\% to 2.37\% in TGARMA(1,1), as $n$ increases from 200 to 1000. The second-regime moving-average coefficient also exhibits appreciable small-sample bias: its mean decreases from 0.2616 to 0.2106, towards the generating value 0.20.

Coverage rates approach the nominal 95\% level as sample size increases. For TGAR(1), coverage rates across all six parameters range from 94.60\% to 95.02\% at \(n=1000\). For TGARMA(1,1), coverage rates for the autoregressive and moving-average coefficients range from 81.27\% to 86.75\% at \(n=200\) and from 92.62\% to 93.40\% at \(n=1000\).

\begin{table}[h]
\centering
\caption{Monte Carlo results for two-regime negative binomial TGAR(1) and TGARMA(1,1) models with log link and GLM-based starting values. Each scenario uses 5,000 replications per sample size. Orders, threshold ($r=1135$, rule $Y_{t-d}>r$) and delay ($d=2$) are known.}
\label{tab:tgar_tgarma_simulation_glm}
\begingroup
\small
\setlength{\tabcolsep}{4pt}
\renewcommand{\arraystretch}{1.02}
\begin{tabular}{rlrrrr@{\hspace{12pt}}rrrr}
\hline
\multicolumn{10}{l}{\textbf{TGAR(1)}} \\
 & & \multicolumn{4}{c}{Regime 1} & \multicolumn{4}{c}{Regime 2} \\
\cline{3-6}\cline{7-10}
$n$ & Measure & $\beta_0$ & $\phi_1$ & $\theta_1$ & $\kappa$ & $\beta_0$ & $\phi_1$ & $\theta_1$ & $\kappa$ \\
 & True value & 7.00 & 0.20 & --- & 179.00 & 7.00 & 0.50 & --- & 427.00 \\
\hline
200 & Mean & 7.0001 & 0.1919 & --- & 185.4156 & 6.9996 & 0.4889 & --- & 480.6085 \\
 & RB (\%) & 0.0007 & -4.0462 & --- & 3.5841 & -0.0061 & -2.2126 & --- & 12.5547 \\
 & Var & $<0.0001$ & 0.0080 & --- & 752.3150 & 0.0001 & 0.0092 & --- & 19109.2552 \\
 & MSE & $<0.0001$ & 0.0081 & --- & 793.3242 & 0.0001 & 0.0093 & --- & 21979.3029 \\
 & CR & 0.9444 & 0.9434 & --- & 0.9364 & 0.9358 & 0.9284 & --- & 0.9332 \\
\hline
500 & Mean & 7.0001 & 0.1977 & --- & 181.9768 & 6.9998 & 0.4960 & --- & 443.6069 \\
 & RB (\%) & 0.0011 & -1.1436 & --- & 1.6630 & -0.0033 & -0.8062 & --- & 3.8892 \\
 & Var & $<0.0001$ & 0.0033 & --- & 282.9259 & $<0.0001$ & 0.0035 & --- & 5097.5949 \\
 & MSE & $<0.0001$ & 0.0033 & --- & 291.7305 & $<0.0001$ & 0.0035 & --- & 5372.3650 \\
 & CR & 0.9526 & 0.9472 & --- & 0.9420 & 0.9448 & 0.9414 & --- & 0.9404 \\
\hline
1000 & Mean & 7.0000 & 0.1998 & --- & 180.5188 & 6.9999 & 0.4984 & --- & 435.7789 \\
 & RB (\%) & -0.0006 & -0.1121 & --- & 0.8485 & -0.0010 & -0.3140 & --- & 2.0559 \\
 & Var & $<0.0001$ & 0.0016 & --- & 133.9734 & $<0.0001$ & 0.0017 & --- & 2357.4555 \\
 & MSE & $<0.0001$ & 0.0016 & --- & 136.2533 & $<0.0001$ & 0.0017 & --- & 2434.0530 \\
 & CR & 0.9498 & 0.9502 & --- & 0.9460 & 0.9472 & 0.9488 & --- & 0.9462 \\
\hline
\multicolumn{10}{l}{\textbf{TGARMA(1,1)}} \\
 & & \multicolumn{4}{c}{Regime 1} & \multicolumn{4}{c}{Regime 2} \\
\cline{3-6}\cline{7-10}
$n$ & Measure & $\beta_0$ & $\phi_1$ & $\theta_1$ & $\kappa$ & $\beta_0$ & $\phi_1$ & $\theta_1$ & $\kappa$ \\
 & True value & 7.00 & 0.20 & 0.15 & 179.00 & 7.00 & 0.50 & 0.20 & 427.00 \\
\hline
200 & Mean & 7.0019 & 0.1896 & 0.1523 & 187.5623 & 7.0012 & 0.4326 & 0.2616 & 490.1458 \\
 & RB (\%) & 0.0271 & -5.1786 & 1.5157 & 4.7834 & 0.0175 & -13.4815 & 30.8059 & 14.7883 \\
 & Var & 0.0002 & 0.0876 & 0.1019 & 807.2029 & 0.0002 & 0.0589 & 0.0655 & 20311.6656 \\
 & MSE & 0.0002 & 0.0877 & 0.1019 & 880.3541 & 0.0002 & 0.0634 & 0.0693 & 24295.0009 \\
 & CR & 0.9297 & 0.8137 & 0.8127 & 0.9297 & 0.8993 & 0.8675 & 0.8577 & 0.9241 \\
\hline
500 & Mean & 7.0009 & 0.1989 & 0.1489 & 182.6175 & 7.0006 & 0.4726 & 0.2254 & 447.0345 \\
 & RB (\%) & 0.0124 & -0.5420 & -0.7534 & 2.0209 & 0.0088 & -5.4719 & 12.7128 & 4.6919 \\
 & Var & $<0.0001$ & 0.0360 & 0.0396 & 296.9128 & $<0.0001$ & 0.0209 & 0.0228 & 4986.8501 \\
 & MSE & $<0.0001$ & 0.0360 & 0.0396 & 309.9395 & $<0.0001$ & 0.0216 & 0.0234 & 5387.2335 \\
 & CR & 0.9464 & 0.8920 & 0.8890 & 0.9388 & 0.9238 & 0.9102 & 0.9066 & 0.9424 \\
\hline
1000 & Mean & 7.0002 & 0.1984 & 0.1512 & 180.8434 & 7.0003 & 0.4890 & 0.2106 & 437.1232 \\
 & RB (\%) & 0.0035 & -0.8245 & 0.8181 & 1.0298 & 0.0037 & -2.2072 & 5.3131 & 2.3708 \\
 & Var & $<0.0001$ & 0.0170 & 0.0183 & 139.5284 & $<0.0001$ & 0.0090 & 0.0100 & 2237.7239 \\
 & MSE & $<0.0001$ & 0.0170 & 0.0183 & 142.8986 & $<0.0001$ & 0.0091 & 0.0101 & 2339.7553 \\
 & CR & 0.9544 & 0.9340 & 0.9262 & 0.9458 & 0.9436 & 0.9316 & 0.9316 & 0.9492 \\
\hline
\end{tabular}
\par\smallskip
\endgroup
\end{table}
\FloatBarrier

\section{Application: Monthly Hospital Admissions in Brazil}
\label{sec:application}

We fit the proposed model to the monthly number of hospital admissions for respiratory causes among adults aged 60 and older in the municipality of São Paulo, Brazil. The data were obtained from the Brazilian Hospital Information System (SIH/SUS), maintained by DATASUS and publicly available at \url{https://datasus.saude.gov.br}. The series covers January 2003 through December 2019, yielding 204 monthly observations. We restrict attention to the pre-2020 period to avoid the structural disruption associated with the COVID-19 pandemic. The monthly frequency motivates the inclusion of annual seasonal components with period 12.

Let $Y_t$ denote the number of admissions in month $t$, and let
$\F_{t-1}$ denote the information available before observing $Y_t$.
We fit a two-regime TGARMA model with a negative binomial
conditional distribution and logarithmic link,
\[
Y_t\mid\F_{t-1},S_t=s
\sim \operatorname{NB}\bigl(\mu_t^{(s)},\kappa^{(s)}\bigr),
\qquad
\log\bigl(\mu_t^{(s)}\bigr)=\eta_t^{(s)},
\]
where
\[
\operatorname{Var}(Y_t\mid\F_{t-1},S_t=s)
=
\mu_t^{(s)}
+\frac{\bigl(\mu_t^{(s)}\bigr)^2}{\kappa^{(s)}}.
\]
The regime indicator is determined by a delayed observation,
\[
S_t=
\begin{cases}
1, & Y_{t-d}\le r,\\
2, & Y_{t-d}>r.
\end{cases}
\]
For each regime $s\in\{1,2\}$, the linear predictor used in the application is
\[
\eta_t^{(s)}
=
\beta_0^{(s)}
+\beta_1^{(s)} z_t
+\sum_{k=1}^{3}
\left[
a_k^{(s)}\sin\left(\frac{2\pi kt}{12}\right)
+b_k^{(s)}\cos\left(\frac{2\pi kt}{12}\right)
\right]
+\phi_1^{(s)}
\left\{
\log(Y_{t-1}^*) -
\mathbf{x}_{t-1}^{\top}\boldsymbol{\beta}^{(S_{t-1})}
\right\},
\]
where $z_t=(t-\bar t)/s_{\mathrm{time}}$ is the standardized time trend, $Y_{t-1}^*=\max(Y_{t-1},0.3)$ is used inside the logarithm, and $\mathbf{x}_t$ contains the intercept, standardized trend, and Fourier seasonal regressors. The moving-average orders were fixed at zero in both regimes. This restriction follows \bcitep{albarracin2019generalized}, who discusses the strong collinearity that may arise when autoregressive and moving-average terms are included simultaneously in GARMA-type predictors.
\begin{figure}[htbp]
\centering
\includegraphics[width=0.9\textwidth]{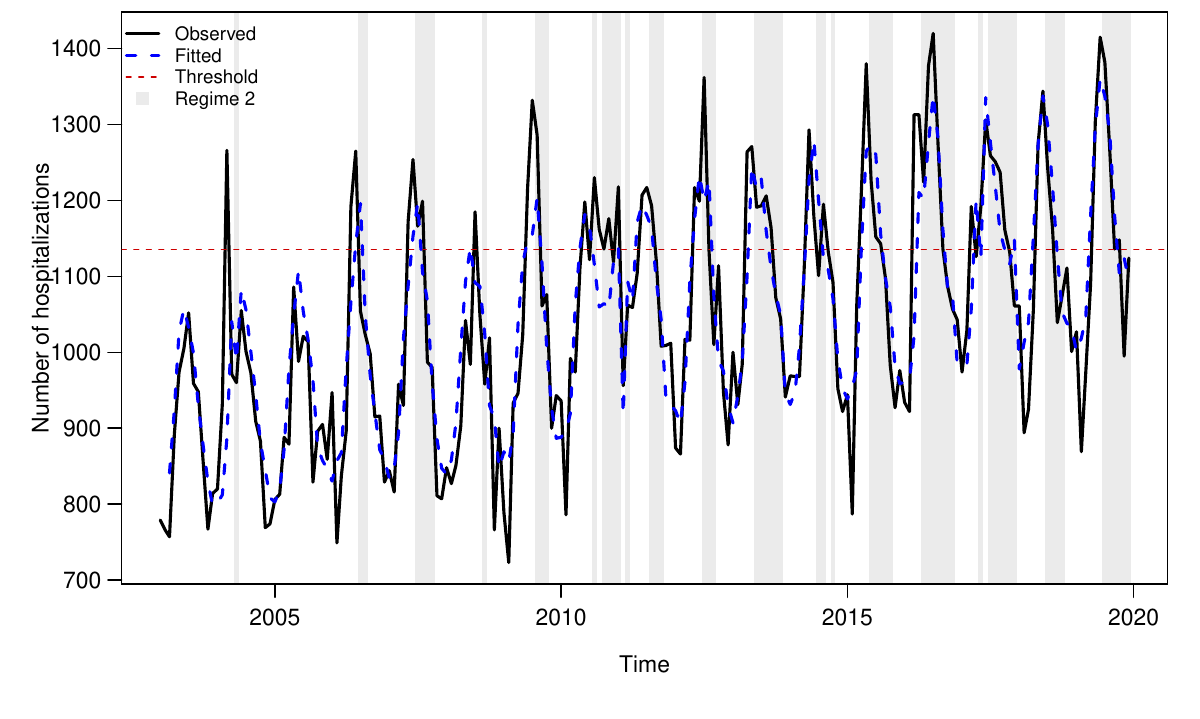}
\caption{Observed and fitted monthly hospitalization counts under the selected threshold GARMA-BN model. The shaded periods correspond to the second estimated regime.}
\label{fig:global_fit}
\end{figure}
\FloatBarrier
Figure~\ref{fig:global_fit} shows that the fitted values follow the main movements observed in the admission series, including its seasonal pattern. The estimated threshold separates the observations into a more frequent lower-count regime and a less frequent higher-count regime. The shaded periods, corresponding to regime 2, occur during months preceded by comparatively high admission counts, which is consistent with the delayed-threshold structure of the fitted model.

For each candidate specification, continuous parameters were estimated by PMLE, conditioning on the initial observations required by the largest lag and on the observed covariate path. The grid considered $(p_1,p_2)\in\{0,1,2,3,4\}^2$, $q_1=q_2=0$, $d\in\{1,2,3,4,5\}$, and thresholds defined by a grid of empirical quantiles of the observed counts.  We imposed a finite-sample trimming rule requiring at least 15\% of the effective observations in each active regime. Model selection used the common likelihood window $t=6,\ldots,204$, giving $n_{\mathrm{sel}}=199$. The retained configuration minimized the PMLE-based BIC criterion among the admissible candidates. After selection, the continuous parameters were reestimated by PMLE on the model-specific window $t=3,\ldots,204$, giving $n_{\mathrm{final}}=202$. Standard errors are based on the inverse observed negative log-likelihood Hessian at the final PMLE and treat the selected orders, delay, threshold and deterministic design as fixed; the coverage qualifications in Subsection~\ref{subsec:asymptotic} apply.
\input{TabelasA/table_model}
The selected model has $(p_1,p_2,q_1,q_2,d)=(1,1,0,0,2)$ and integer cut $r=1135$ (the quantile-grid value was 1135.10). Thus the regime for month $t$ depends on whether $Y_{t-2}>1135$. The regime sizes in the final PMLE window are 141 and 61. There are 20 continuous parameters: eight regression coefficients, one AR coefficient, and one NB shape parameter in each regime. The final PMLE gives log-likelihood $-1159.7474$, AIC $2359.4948$ and BIC $2425.6601$, calculated with 202 observations (Table~\ref{tab:model_specification}). Numerical convergence and a positive-definite observed negative log-likelihood Hessian were recorded; interval coverage after structural selection has not been established.
\input{TabelasA/parameters}

Table~\ref{tab:parameters} reports the estimates. The recorded trend coefficients are positive in both regimes. Converting only the explicit standardized-trend term gives monthly multiplicative changes of approximately 0.131\% and 0.098\%, or 1.58\% and 1.18\% over twelve months, holding the remaining predictor contributions fixed. These are transformations of the regression component, not marginal growth rates of the full threshold process. The fitted seasonal coefficients differ across regimes, but individual within-regime tests do not establish between-regime differences, and their nominal significance does not account for selection. The AR estimates, $\hat\phi^{(1)}=0.2175$ and $\hat\phi^{(2)}=0.4649$, describe a larger fitted coefficient in the second regime; this is not a calibrated test of different persistence. Both satisfy the AR(1) diagnostic $|\hat\phi^{(s)}|<1$, which does not prove global stationarity. The shape estimates are $\hat\kappa^{(1)}=179.24$ and $\hat\kappa^{(2)}=427.04$. At equal conditional means, the larger $\kappa^{(s)}$ implies less overdispersion; across observations, the variance-to-mean ratio is $1+\mu_t^{(s)}/\kappa^{(s)}$, so the means must also be considered.

\begin{figure}[htbp]
\centering
\includegraphics[width=0.9\textwidth]{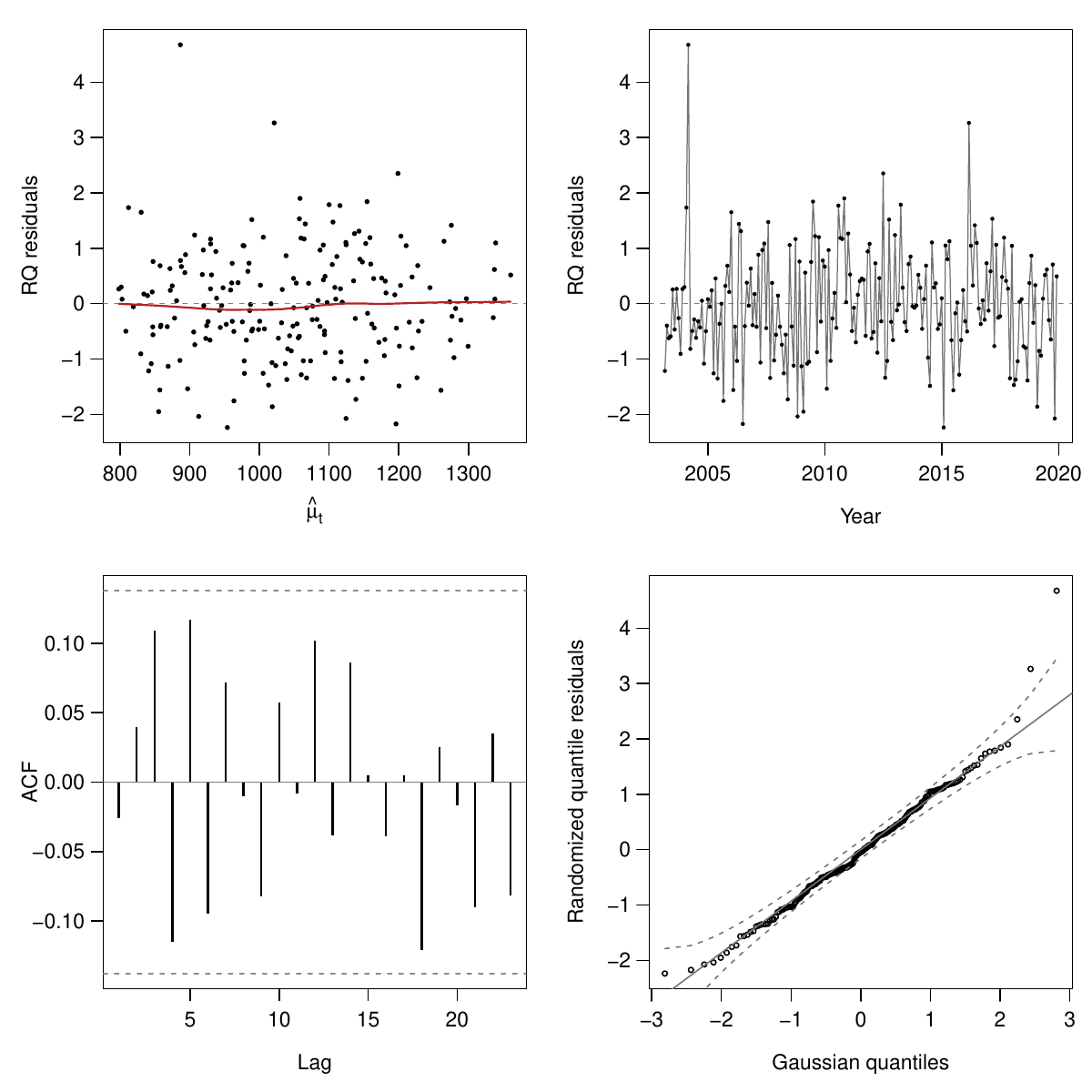}
\caption{Global diagnostic plots based on randomized quantile residuals for the selected threshold GARMA-BN model. The panel shows residuals versus fitted values, residuals versus time, the sample autocorrelation function, and the normal Q--Q plot with reference envelope.}
\label{fig:global_diagnostics}
\end{figure}

Model adequacy was assessed using randomized quantile residuals, which are suitable for discrete response models and permit standard checks of residual serial dependence and distributional agreement. The Ljung--Box test applied to these residuals gives $p=0.166$, providing no strong evidence of remaining serial correlation at the lags considered. The residual plots in Figure~\ref{fig:global_diagnostics} are consistent with this result: the residuals are centered around zero and the sample autocorrelations remain within the reference bands for most lags. The Shapiro--Wilk test rejects exact normality of the randomized quantile residuals ($p=0.000650$), and the Q--Q plot shows some tail departure. This suggests caution in interpreting the normal approximation in the extremes. 
To assess the empirical contribution of the threshold mechanism, we also fitted a non-threshold GARMA-BN benchmark with the same Negative Binomial conditional distribution. Table~\ref{tab:tgarma_vs_garma} compares the selected threshold model with this benchmark. The threshold model improves the maximized log-likelihood, AIC, RMSE, MAE, and descriptive $R^2$, indicating a better in-sample representation of the admission series. The BIC favors the GARMA-BN benchmark, reflecting the stronger penalty imposed on the additional regime-specific parameters in the threshold model. Thus, while the non-threshold benchmark remains competitive under BIC, the threshold specification offers a structured description of how level, seasonality, dispersion, and persistence vary between lower- and higher-count periods.
\FloatBarrier
\input{Tabelas/table_tgarma_vs_garma}
\FloatBarrier

\section{Conclusion}
\label{sec:conclusion}
We introduced a threshold generalization of GARMA models for non-Gaussian time series in which the active regression, autoregressive, moving-average, and, when appropriate, dispersion components may vary across regimes determined by a lagged threshold variable. The proposed formulation retains the observation-driven structure of GARMA models while incorporating an interpretable mechanism for regime switching. From an inferential standpoint, the paper develops a profile-likelihood approach for threshold selection and PMLE for the continuous parameter vector, including the score vector, conditional information matrix, observed Hessian and OPG matrix. This distinction is important: the regular asymptotic theory applies to the continuous parameters, whereas the threshold parameters require profile-likelihood treatment and follow a non-standard asymptotic regime, in line with the classical threshold literature \bcitep{chan1990likelihood,chan1993consistency,hansen2011threshold}.

The empirical application to monthly hospital admissions in S\~ao Paulo illustrates that the threshold specification can capture regime-specific differences in persistence, seasonal structure, and dispersion that are not available in a single-regime GARMA benchmark.

\newpage
\appendix

\section{Conditional Exponential-Family Components}
\label{app:exp_family_components}

This appendix records the distributional quantities used in the likelihood
notation of Section~\ref{sec:proposed_model}. For a fixed regime $s$, the
conditional law of $Y_t$ given $\F_{t-1}$ and $S_t=s$ is written in the
exponential-dispersion form
\[
f(y_t\mid\F_{t-1},S_t=s)
=
\exp\left\{
\frac{y_t\omega_t^{(s)}-b(\omega_t^{(s)})}{\varphi^{(s)}}
+c(y_t,\varphi^{(s)})
\right\}.
\]
The conditional mean and variance satisfy
\[
\mu_t^{(s)}=b'\{\omega_t^{(s)}\},
\qquad
\operatorname{Var}(Y_t\mid\F_{t-1},S_t=s)
=
\varphi^{(s)}b''\{\omega_t^{(s)}\}
=
\varphi^{(s)}V(\mu_t^{(s)}).
\]

For the Poisson, Binomial, Gamma, Normal and Inverse Gaussian cases, the
functions $b(\cdot)$, $c(\cdot,\cdot)$ and $V(\cdot)$ are the standard GLM
quantities. Conditional families with additional shape or precision
parameters, such as the Negative Binomial parametrization used in the empirical
application, are handled by writing the corresponding conditional
log-likelihood directly and by deriving the associated score and information
blocks from that parametrization.

\section{Proof of Lemma~\ref{lem:conditional_mean_score}}

The following identities are used in the construction of the conditional
information matrix.

\begin{lemma}[Conditional score-information identities]
\label{lem:conditional_mean_score}
Let $(\mathbf r,d)$ be fixed and admissible, and suppose that the conditional
exponential-dispersion family is evaluated at an interior point of the mean
domain, with parameter-independent support, $V(\mu)>0$, differentiable
variance function $V$, and differentiable mean recursion with respect to
$\boldsymbol\psi$. For $t=m+1,\ldots,T$, let $\boldsymbol\delta_t$ denote the
transpose of the row of $\Delta$ associated with time $t$. Then the following
identities hold:
\begin{enumerate}
\item[(i)]
\[
\frac{\partial\ell_t}{\partial\mu_t^{(S_t)}}
=
\frac{Y_t-\mu_t^{(S_t)}}{\varphi^{(S_t)}V_t^{(S_t)}},
\qquad
\mathbb E_{\boldsymbol\theta}
\left(
\frac{\partial\ell_t}{\partial\mu_t^{(S_t)}}
\mid\F_{t-1}
\right)
=0.
\]
\item[(ii)]
\[
U_t^{(\psi)}(\boldsymbol\theta;\mathbf r,d)
=
\boldsymbol\delta_t
\frac{D_t^{(S_t)}}{\varphi^{(S_t)}V_t^{(S_t)}}
(Y_t-\mu_t^{(S_t)}).
\]
\item[(iii)]
\[
\mathbb E_{\boldsymbol\theta}
\left[
U_t^{(\psi)}U_t^{(\psi)\top}
\mid\F_{t-1}
\right]
=
\mathbb E_{\boldsymbol\theta}
\left[
-
\frac{\partial^2\ell_t}
{\partial\boldsymbol\psi\,\partial\boldsymbol\psi^\top}
\mid\F_{t-1}
\right]
=
\frac{\{D_t^{(S_t)}\}^2}{\varphi^{(S_t)}V_t^{(S_t)}}
\boldsymbol\delta_t\boldsymbol\delta_t^\top.
\]
\end{enumerate}
\end{lemma}

\begin{proof}
Fix $t$ and write $s=S_t(\mathbf r,d)$. Since the threshold configuration is
fixed and admissible, $S_t$ is $\F_{t-1}$-measurable. Thus, conditional on
$\F_{t-1}$, the active regime $s$, the conditional mean $\mu_t^{(s)}$, the
dispersion $\varphi^{(s)}$, and the variance function value
$V_t^{(s)}=V(\mu_t^{(s)})$ are fixed. The active-regime log-likelihood
contribution is
\[
\ell_t
=
\frac{Y_t\omega_t^{(s)}-b\{\omega_t^{(s)}\}}{\varphi^{(s)}}
+c(Y_t,\varphi^{(s)}),
\qquad
\mu_t^{(s)}=b'\{\omega_t^{(s)}\}.
\]
Since
$d\mu_t^{(s)}/d\omega_t^{(s)}
=b''\{\omega_t^{(s)}\}=V(\mu_t^{(s)})>0$, it follows that
$d\omega_t^{(s)}/d\mu_t^{(s)}=1/V_t^{(s)}$. Hence
\[
\frac{\partial\ell_t}{\partial\mu_t^{(s)}}
=
\frac{\partial\ell_t}{\partial\omega_t^{(s)}}
\frac{\partial\omega_t^{(s)}}{\partial\mu_t^{(s)}}
=
\frac{Y_t-b'\{\omega_t^{(s)}\}}{\varphi^{(s)}}
\frac{1}{V_t^{(s)}}
=
\frac{Y_t-\mu_t^{(s)}}{\varphi^{(s)}V_t^{(s)}}.
\]
Taking conditional expectation and using
$\mathbb E_{\boldsymbol\theta}(Y_t\mid\F_{t-1})=\mu_t^{(s)}$ proves (i).

For (ii), use the chain rule and the definition
$D_t^{(s)}=\partial\mu_t^{(s)}/\partial\eta_t^{(s)}$. Since
$\boldsymbol\delta_t=\partial\eta_t^{(s)}/\partial\boldsymbol\psi$, we have
\[
\frac{\partial\mu_t^{(s)}}{\partial\boldsymbol\psi}
=
D_t^{(s)}\boldsymbol\delta_t.
\]
Therefore,
\[
U_t^{(\psi)}
=
\frac{\partial\ell_t}{\partial\mu_t^{(s)}}
\frac{\partial\mu_t^{(s)}}{\partial\boldsymbol\psi}
=
\boldsymbol\delta_t
\frac{D_t^{(s)}}{\varphi^{(s)}V_t^{(s)}}
(Y_t-\mu_t^{(s)}),
\]
which is (ii) after replacing $s$ by $S_t$.

For (iii), first use (ii) and the conditional variance formula from the model
specification:
\[
\mathbb E_{\boldsymbol\theta}
\left[
U_t^{(\psi)}U_t^{(\psi)\top}
\mid\F_{t-1}
\right]
=
\frac{\{D_t^{(s)}\}^2}{\{\varphi^{(s)}V_t^{(s)}\}^2}
\boldsymbol\delta_t\boldsymbol\delta_t^\top
\mathbb E_{\boldsymbol\theta}
\left[
(Y_t-\mu_t^{(s)})^2
\mid\F_{t-1}
\right]
=
\frac{\{D_t^{(s)}\}^2}{\varphi^{(s)}V_t^{(s)}}
\boldsymbol\delta_t\boldsymbol\delta_t^\top.
\]
It remains to verify the negative expected Hessian representation. Applying
the chain rule twice,
\[
\frac{\partial^2\ell_t}{\partial\boldsymbol\psi\,\partial\boldsymbol\psi^\top}
=
\frac{\partial^2\ell_t}{\partial\{\mu_t^{(s)}\}^2}
\frac{\partial\mu_t^{(s)}}{\partial\boldsymbol\psi}
\frac{\partial\mu_t^{(s)}}{\partial\boldsymbol\psi^\top}
+
\frac{\partial\ell_t}{\partial\mu_t^{(s)}}
\frac{\partial^2\mu_t^{(s)}}{\partial\boldsymbol\psi\,\partial\boldsymbol\psi^\top}.
\]
The second term has conditional expectation zero by (i), because
$\partial^2\mu_t^{(s)}/\partial\boldsymbol\psi\,\partial\boldsymbol\psi^\top$
is $\F_{t-1}$-measurable for the fixed threshold configuration. Moreover,
\[
\frac{\partial^2\ell_t}{\partial\{\mu_t^{(s)}\}^2}
=
-\frac{1}{\varphi^{(s)}V_t^{(s)}}
-\frac{\{Y_t-\mu_t^{(s)}\}V'(\mu_t^{(s)})}
{\varphi^{(s)}\{V_t^{(s)}\}^2},
\]
so
\[
-
\mathbb E_{\boldsymbol\theta}
\left(
\frac{\partial^2\ell_t}{\partial\{\mu_t^{(s)}\}^2}
\mid\F_{t-1}
\right)
=
\frac{1}{\varphi^{(s)}V_t^{(s)}}.
\]
Substituting
$\partial\mu_t^{(s)}/\partial\boldsymbol\psi
=D_t^{(s)}\boldsymbol\delta_t$ gives the same matrix as the conditional
second moment. Replacing $s$ by $S_t$ proves (iii).
\end{proof}

\section{Proof of Theorem~\ref{thm:continuous_pmle}}
\label{app:continuous_pmle_proof}

The argument is written under the probability measure
$\mathbb P_{\boldsymbol\theta_0}$ and for a fixed admissible threshold
configuration $(\mathbf r,d)$. Conditioning on the initial values used to start
the recursion, (A1) implies that $S_t(\mathbf r,d)$ is
$\F_{t-1}$-measurable. Thus the likelihood contributions are sequential
conditional likelihood contributions for the continuous parameter vector.

Under (A3), differentiation may be interchanged with conditional integration.
Since each conditional density or probability mass function integrates or sums
to one over a parameter-independent support,
\[
\mathbb E_{\boldsymbol\theta_0}
\{U_t(\boldsymbol\theta_0;\mathbf r,d)\mid\F_{t-1}\}
=\mathbf 0 .
\]
Consequently, $M_T=\sum_{t=m+1}^{T}U_t(\boldsymbol\theta_0;\mathbf r,d)$ is a
square-integrable vector martingale with respect to $\{\F_t\}$, and its
predictable quadratic variation is
\[
\langle M\rangle_T
=
\sum_{t=m+1}^{T}
\mathbb E_{\boldsymbol\theta_0}
\left[
U_t(\boldsymbol\theta_0;\mathbf r,d)
U_t(\boldsymbol\theta_0;\mathbf r,d)^\top
\mid\F_{t-1}
\right]
=
\mathcal K_T(\boldsymbol\theta_0;\mathbf r,d).
\]
For every fixed vector $a$, the scalar array
$a^\top U_t(\boldsymbol\theta_0;\mathbf r,d)$ satisfies the conditional
Lindeberg condition by (A7), and its normalized predictable variation converges
in probability to
$a^\top\mathcal I(\boldsymbol\theta_0;\mathbf r,d)a$. The scalar martingale
central limit theorem and the Cramer--Wold device therefore yield
\[
\frac{1}{\sqrt{n^\ast}}U_T(\boldsymbol\theta_0;\mathbf r,d)
\xrightarrow{d}
\mathcal N\{\mathbf 0,\mathcal I(\boldsymbol\theta_0;\mathbf r,d)\}.
\]

For consistency, define the normalized likelihood contrast
\[
Q_T(\boldsymbol\theta;\mathbf r,d)
=
\frac{1}{n^\ast}
\{\ell_T(\boldsymbol\theta;\mathbf r,d)
-\ell_T(\boldsymbol\theta_0;\mathbf r,d)\}.
\]
At the population level, correct conditional specification gives the
conditional Kullback--Leibler contrast
\[
\mathbb E_{\boldsymbol\theta_0}
\{\ell_t(\boldsymbol\theta;\mathbf r,d)
-\ell_t(\boldsymbol\theta_0;\mathbf r,d)\mid\F_{t-1}\}
\leq 0,
\]
with equality only when the two conditional distributions coincide
$\mathbb P_{\boldsymbol\theta_0}$-almost surely. By (A5), this can occur only
at $\boldsymbol\theta=\boldsymbol\theta_0$. Hence the limiting criterion is
uniquely maximized at $\boldsymbol\theta_0$. Assumptions (A4) and (A6) make the
effect of the initial conditioning values asymptotically negligible and give
the required uniform convergence of $Q_T$ in a neighbourhood of
$\boldsymbol\theta_0$. The standard local argmax theorem for likelihood
estimators then implies that any interior sequence of local maximizers in that
neighbourhood is consistent.

Since $\boldsymbol\theta_0$ is an interior point by (A2), the preceding
consistency argument applies on a neighbourhood contained in $\Theta$. Together
with the nonsingularity of the limiting information matrix, the local regularity
conditions in (A3), (A6) and (A7) imply the existence, with probability tending
to one, of a locally unique interior maximizer in a neighbourhood of
$\boldsymbol\theta_0$. The following expansion is therefore applied to any such
sequence of local maximizers.

It remains to derive the limiting distribution. For an interior local maximizer,
\[
U_T\{\widehat{\boldsymbol\theta}(\mathbf r,d);\mathbf r,d\}
=\mathbf 0.
\]
Let $h_T=\widehat{\boldsymbol\theta}(\mathbf r,d)-\boldsymbol\theta_0$ and
define
\[
\bar{\mathcal H}_T
=
\int_0^1
\mathcal H_T\{\boldsymbol\theta_0+a h_T;\mathbf r,d\}\,da .
\]
The integral form of Taylor's theorem applied to the score vector gives
\[
\mathbf 0
=
U_T(\boldsymbol\theta_0;\mathbf r,d)
-
\bar{\mathcal H}_T h_T .
\]
Therefore,
\[
\sqrt{n^\ast}
\{\widehat{\boldsymbol\theta}(\mathbf r,d)-\boldsymbol\theta_0\}
=
\left[
\frac{1}{n^\ast}\bar{\mathcal H}_T
\right]^{-1}
\frac{1}{\sqrt{n^\ast}}
U_T(\boldsymbol\theta_0;\mathbf r,d).
\]
By consistency and the uniform Hessian convergence in (A6),
\[
\frac{1}{n^\ast}\bar{\mathcal H}_T
\xrightarrow{p}
\mathcal I(\boldsymbol\theta_0;\mathbf r,d).
\]
The positive definiteness in (A7) implies that the limiting matrix is
nonsingular. Slutsky's theorem then gives
\[
\sqrt{n^\ast}
\{\widehat{\boldsymbol\theta}(\mathbf r,d)-\boldsymbol\theta_0\}
\xrightarrow{d}
\mathcal N\left(
\mathbf 0,
\mathcal I(\boldsymbol\theta_0;\mathbf r,d)^{-1}
\right).
\]
This proves the result for the regular continuous component. The argument does
not establish a regular limiting distribution for
$(\widehat{\mathbf r},\widehat d)$, because threshold parameters enter the
likelihood through discontinuous regime indicators.

\bibliographystyle{apalike}
\bibliography{bibliografia.bib}


\end{document}

%% file: TabelasA/table_model.tex
\begin{table}[h]
\centering
\small
\caption{Selected threshold GARMA-BN specification.}
\label{tab:model_specification}
\resizebox{\textwidth}{!}{%
\begin{tabular}{lllrrrrrrr}
\toprule
Model & $(p_1,p_2,q_1,q_2,d,r)$ & Window & $n$ & $k$ & LogLik & AIC & BIC & $n_1$ & $n_2$ \\
\midrule
TGARMA-BN & (1,1,0,0,2,1135.10) & 3--204 & 202 & 20 & -1159.7474 & 2359.4948 & 2425.6601 & 141 & 61 \\
\bottomrule
\end{tabular}}
\par\medskip\begin{minipage}{\textwidth}\footnotesize
Selection used observations 6--204 ($n_{\mathrm{sel}}=199$); final results use the window shown above. Regime sizes exclude the conditioning observations.
\end{minipage}
\end{table}

%% file: TabelasA/parameters.tex
\begin{table}[htbp]
\centering
\small
\caption{PMLE estimates and conditional Wald intervals for the selected TGARMA-BN model.}
\label{tab:parameters}
\resizebox{\textwidth}{!}{%
\begin{tabular}{lrrrr}
\toprule
 & \multicolumn{2}{c}{Regime 1}
 & \multicolumn{2}{c}{Regime 2} \\
\cmidrule(lr){2-3}\cmidrule(lr){4-5}
Parameter & Est. (SE) & 95\% CI & Est. (SE) & 95\% CI \\
\midrule
$\beta_0$
& $6.9350^{***}\ (0.0095)$ & $[6.9163,\ 6.9537]$
& $6.9882^{***}\ (0.0186)$ & $[6.9518,\ 7.0246]$ \\
$\beta_1$
& $0.0772^{***}\ (0.0089)$ & $[0.0597,\ 0.0947]$
& $0.0578^{***}\ (0.0140)$ & $[0.0303,\ 0.0853]$ \\
$a_1$
& $-0.0416^{**}\ (0.0134)$ & $[-0.0680,\ -0.0153]$
& $-0.0061\ (0.0236)$ & $[-0.0524,\ 0.0403]$ \\
$b_1$
& $-0.1447^{***}\ (0.0115)$ & $[-0.1672,\ -0.1221]$
& $-0.0397^{*}\ (0.0195)$ & $[-0.0779,\ -0.0015]$ \\
$a_2$
& $-0.0096\ (0.0114)$ & $[-0.0320,\ 0.0127]$
& $0.0589^{**}\ (0.0197)$ & $[0.0202,\ 0.0976]$ \\
$b_2$
& $0.0211\ (0.0108)$ & $[-0.0001,\ 0.0423]$
& $0.0154\ (0.0190)$ & $[-0.0218,\ 0.0525]$ \\
$a_3$
& $0.0082\ (0.0095)$ & $[-0.0105,\ 0.0269]$
& $-0.0037\ (0.0126)$ & $[-0.0284,\ 0.0209]$ \\
$b_3$
& $0.0028\ (0.0100)$ & $[-0.0167,\ 0.0223]$
& $-0.0126\ (0.0138)$ & $[-0.0397,\ 0.0146]$ \\
$\phi_1$
& $0.2175^{*}\ (0.0879)$ & $[0.0451,\ 0.3898]$
& $0.4649^{***}\ (0.1236)$ & $[0.2225,\ 0.7072]$ \\
$\kappa$
& $179.24\ (25.28)$ & $[135.95,\ 236.33]$
& $427.04\ (106.91)$ & $[261.44,\ 697.55]$ \\
\bottomrule
\end{tabular}}

\par\medskip
\begin{minipage}{\textwidth}
\footnotesize
Regime 1 corresponds to $Y_{t-2}\le 1135.10$ and
Regime 2 to $Y_{t-2}>1135.10$. The trend is standardized.
Annual, semiannual, and third-harmonic terms correspond
to Fourier sine/cosine pairs with periods 12, 6, and
4 months, respectively. For $\kappa$, standard errors
use the delta method and confidence intervals are obtained
by exponentiating the log-scale Wald limits.
$^{*}p<0.05$, $^{**}p<0.01$, $^{***}p<0.001$
for the regression and autoregressive coefficients.
\end{minipage}
\end{table}

%% file: Tabelas/table_tgarma_vs_garma.tex
\begin{table}[ht]
\centering
\small
\caption{In-sample comparison between the selected threshold GARMA-BN model and the non-threshold GARMA-BN benchmark.}
\label{tab:tgarma_vs_garma}
\resizebox{\textwidth}{!}{%
\begin{tabular}{lrrrrrrrr}
\toprule
Model & $n$ & $k$ & LogLik & AIC & BIC & RMSE & MAE & $R^2$ \\
\midrule
TGARMA-BN & 202 & 20 & -1159.7474 & 2359.4948 & 2425.6601 & 77.0302 & 59.3653 & 0.7582 \\
GARMA-BN & 202 & 10 & -1176.8578 & 2373.7157 & 2406.7984 & 82.0311 & 64.4012 & 0.7258 \\
\bottomrule
\end{tabular}}
\end{table}